\documentclass[prb,superscriptaddress,amsmath,amssymb,twocolumn]{revtex4-2}
\pdfoutput=1
\usepackage{physics}
\usepackage{lipsum}
\usepackage{soul}
\usepackage{booktabs}
\usepackage{natbib}
\usepackage{float}
 \usepackage{subfloat}
\usepackage[english]{babel}
\makeatother
\usepackage{layouts} 
\usepackage{graphicx}
\usepackage{caption}
\usepackage{subcaption}
\usepackage[usenames,dvipsnames]{color}
\usepackage{stackengine}
\usepackage[T1]{fontenc}
\usepackage[colorlinks=true,allcolors=blue]{hyperref}

\newcommand{\Dadd}[1]{\textcolor{Red}{#1}}

\begin{document}

\title{Topological phases and spontaneous symmetry breaking: the revenge of the original Su-Schrieffer-Heeger model}

\author{Polina Matveeva}
\affiliation{Department of Physics, Bar-Ilan University, Ramat Gan, 52900, Israel}
\author{Dmitri Gutman}
\affiliation{Department of Physics, Bar-Ilan University, Ramat Gan, 52900, Israel}
\author{Sam T.~Carr}
\affiliation{School of Engineering,  Mathematics and Physics,  University of Kent, Canterbury CT2 7NH, United Kingdom}

\begin{abstract}
We study the interplay of spontaneous symmetry breaking and topological properties in interacting one-dimensional models.  We solve these models using bozonization and identify topologically non-trivial phases by counting the additional degeneracy (affiliated with the edge modes) of a finite-size system relative to the infinite one.   We find even if the mean-field solution is topological,  this may not be true when it arises from spontaneous symmetry breaking,  including in the Su-Schrieffer-Heeger (SSH) model. This implies that the original SSH model,  as presented by Su, Schrieffer,  and Heeger,  is topologically trivial,  as opposed to its mean-field version.  A spinful version,  on the other hand,  does exhibit a topologically non-trivial phase.  In that state,  both mean-field solutions are topologically non-trivial and correspond to non-interacting SSH chains in the opposite phases with the winding number $\nu=1$.   
 We show that this phase is protected by a chiral symmetry,   similar to the non-interacting phases.  

\end{abstract}

\maketitle
\section{Introduction} 

Topological phases of matter represent states that feature global physical properties that are robust with respect to perturbations.  The first famous example of a two-dimensional topological phase is the Quantum Hall effect \cite{Klitzing1980, Thouless1982}.  Later on,  this notion was extended to symmetry-protected topological phases.  These are phases that are robust to perturbation so long as certain symmetries,  such as particle-hole, time-reversal,  and/or chiral are preserved.   Such phases have been fully classified and well-understood \cite{Zirnbauer1996,Altland1997,Kitaev2009,Ryu2010} for non-interacting electrons.  This has been extended to weakly-interacting particles \cite{Morimoto2015, Fidkowski2010,Fidkowski2011},  although a full general classification remains an open question.  

While the general classification of interacting topological phases in an arbitrary dimension is incomplete,  one-dimensional models represent a special case where the symmetry classification is established \cite{Turner2011,Wen2009}.  While such a classification exists,  it is quite abstract,  and therefore it is useful to look at specific models elucidating the physical nature of different possible phases.     

The simplest example of a non-interacting one-dimensional topological insulator is the Su-Schrieffer-Heeger (SSH) chain \cite{SSH1979,Kivelson1988}, which has two thermodynamically equivalent phases.  One is topological and exhibits edge states,  while the other is trivial.   There have been many studies of this model and its extensions,  including the multi-band and interacting ones \cite{Sirker2014, Li2014,  DeGottardi2018,Yahyavi_2018,Zegarra2019,Nersesyan2020, Giamarchi2023,Matveeva2023,Melo2023,Matveeva2024}.  There is also a superconducting version known as the Kitaev chain \cite{Kitaev2001},  which features Majorana edge modes and represents a platform for realizing topological qubits \cite{Microsoft2025,Alicea2011}. 

While the SSH model is now used almost exclusively to refer to a non-interacting model,  exhibiting topological properties,  this is not what the original papers of Su, Schrieffer,  and Heeger \cite{SSH1979} and later Kivelson \cite{Kivelson1988} focused on.  These papers were looking at the properties of polyacetylene molecules and consisted of electrons coupled to phonons.   The dimerization pattern of the SSH model is a property of the mean-field solution of this model (Peierls instability),  and the original papers concentrated on how the solitons between the two distinct mean-field solutions  contributed to the properties of polyacetylene.  To the best of our knowledge, the first discussion of the topological properties of the (non-interacting)  mean-field solutions was by Ryu and Hatsugai \cite{Ryu2006} in 2006,  which kick-started the field.  However,  one can ask the question whether the original SSH model with electron-phonon coupling has non-trivial topological properties? 

In this paper,  we study this question of the interplay between spontaneous symmetry breaking and topology.  Rather than including phonons,  we look at toy lattice models with short-range electron-electron interactions that have the same symmetry-breaking phases as the SSH chain.  We do this first for a spinless chain in Section \ref{single_section} and later generalize to the spinful case in Section  \ref{Section_Model}.  In the spinless case,  we find that there are no edge states,  despite the mean-field solution being the SSH model.  In the spinful case,  we do find an example of a model that exhibits symmetry breaking with topologically protected edge states.  In Section 
\ref{protection_section} we identify the symmetries protecting these edge states and show they are the same as for the non-interacting mean-field solutions.

\begin{figure*}
\begin{subfigure}{.4\linewidth}
\includegraphics[scale=0.11]{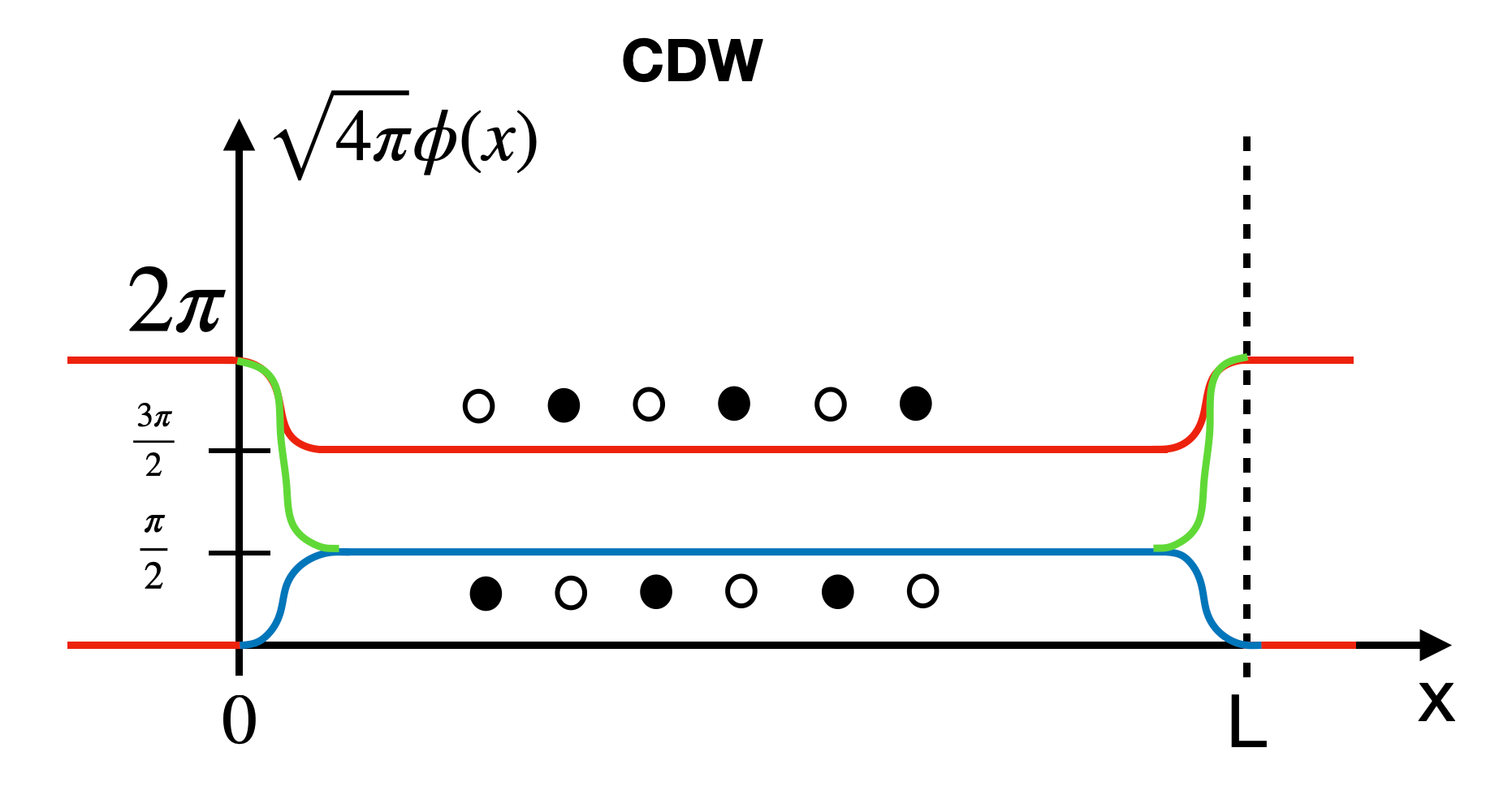}
\caption{}
\label{single_chain_phases_a}
\end{subfigure}
\hspace{2cm}
\begin{subfigure}{.4\linewidth}
\includegraphics[scale=0.11]{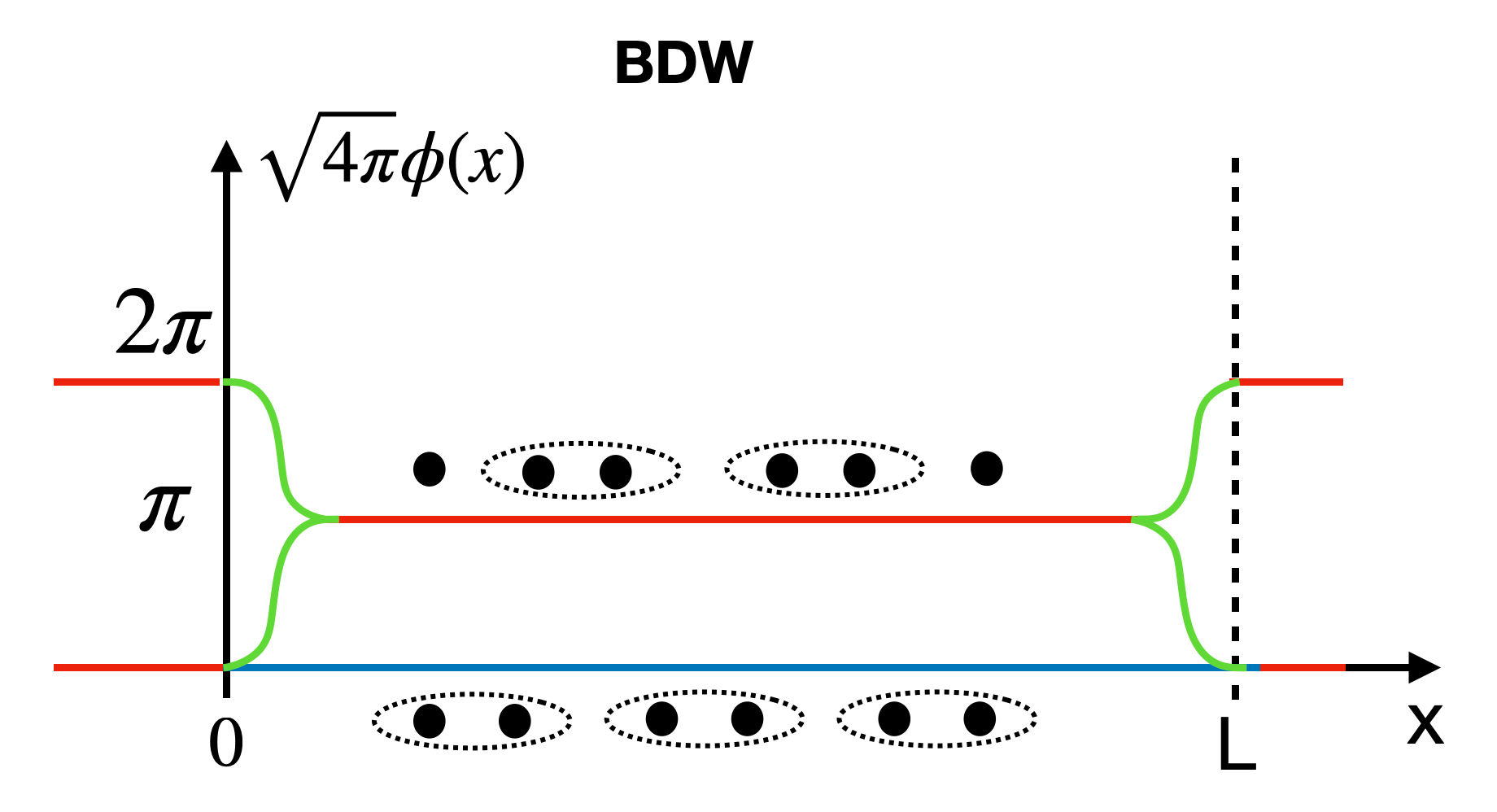}
\caption{}
\label{single_chain_phases_b}
\end{subfigure}
\caption{Profiles of the bosonic fields in a spinless interacting model (\ref{Hsingle}).   The bosonic field $\phi$ is defined on a cylinder such that $\sqrt{4\pi} \phi \rightarrow \sqrt{4\pi} \phi + 2\pi$.  The red and blue solid lines illustrate distinct ground state values and green solid lines show excited states.   a)  charge-density wave phase b) bond-density wave phase }  
\label{single_chain_phases}
\end{figure*}
\section{Topological properties of a spinless chain}
\label{single_section}
\subsection{The model}
We begin by studying a spinless tight-binding problem motivated by the original SSH paper.  In this paper,  the chain was coupled to phonons,  giving rise to the dimerized phases.  However,  we find that by including nearest-neighbor and next-nearest neighbor interactions,  the same physics can be obtained in a model without phonons.  To be specific,  the Hamiltonian is
\begin{align}
\label{single_chain}
H=-t \sum\limits_j \left(c^{\dagger}_{j} c_{j+1} + \text{h.c.}\right) + V_0 \sum\limits_j n_jn_{j+1} + \nonumber \\ 
+ V' \sum\limits_j n_jn_{j+2},
\end{align}
where $c_j$ and $c^{\dagger}_j$ are standard fermionic creation and annihilation operators.  We focus on half-filling. 
 In the absence of $V'$,  the phase diagram of this model is well-understood \cite{gogolin2004bosonization},  for strong repulsive interaction $V_0>t$,  the ground state is an ordered charge-density wave phase that spontaneously breaks $\mathbb{Z}_2$ symmetry.  For a smaller repulsive $V_0$,  the ground state remains gapless,  characterized by power-law correlations (Luttinger liquid).  As we now show,  the presence of the next-nearest neighbor interaction $V' \neq 0$ allows a further broken symmetry ground state for $V'>V_0$, which is of SSH type.   
 
We analyze this model using bosonization.   
Details of the calculation are outlined in Appendix  \ref{bosonization_conventions},  leading to the following bosonic model: 
\begin{align}
\label{Hsingle}
H= \frac{v}{2} \left[K \Pi^2 +K^{-1} (\partial_x \phi)^2\right] + \frac{g}{2(\pi a)^2}\cos[2\sqrt{4\pi} \phi].
\end{align}
For weak interactions,  the Luttinger parameter $K$ is related to the forward scattering amplitude,  $K=1-2a_0(V_0+V')/\pi v_F$,  and $g$ is determined by the Umklapp scattering $g=V_0-V'$.  The Fermi velocity is given by $v_F=2|t|a_0$,  and $a_0$ is the lattice constant.  
For stronger interactions,  there is no general analytic expression relating the fermionic microscopic parameters to the bosonic ones. 
However,  the principle of universality allows us to assume such a relation exists and can be found by numerics,  if this detail is required.  

For Luttinger parameter $K<1/2$,  the Umklapp term  is relevant (in the renormalization group sense),  which leads to a gap opening.  The nature of the ground state is determined by the sign of $g$.  
\subsection{Order parameters}

For $g>0$,  the ground state is a charge-density wave with the order parameter: 
\begin{align}
\label{O_CDW}
\mathcal{O}_{\text{CDW}}=  (-1)^j c^{\dagger}_{j} c_{j}  + \text{h.c.} \propto  \sin[\sqrt{4\pi} \phi],
\end{align}
while for $g<0$ the ground state is a bond-density wave,  which we also call the SSH phase in order to emphasize that it has a similar dimerization as an SSH chain:
\begin{align}
\label{O_BDW}
\mathcal{O}_{\text{SSH}}=  (-1)^j c^{\dagger}_{j} c_{j+1}  + \text{h.c.} \propto   \cos[\sqrt{4\pi} \phi].
\end{align} 
This is easily derived by a semi-classical analysis of the bosonic Hamiltonian  \eqref{Hsingle} by looking at the values of the field $\phi$ that minimizes the cosine potential.  We then see which order parameter is non-zero for this value of $\phi$.  This gives: 
\begin{align}
& g<0  \rightarrow   \sqrt{4\pi} \phi= 0 \mod \pi \hspace{0.1cm} \rightarrow \text{ SSH (bond-density wave) }  \nonumber \\
& g>0   \rightarrow  \sqrt{4\pi} \phi= \pi/2 \mod \pi \hspace{0.1cm} \rightarrow \text{Charge-density wave}, \nonumber
\end{align}
as stated previously.  Going beyond semi-classical analysis does not change this result,  so long as the cosine term is relevant \cite{Zamolodchikov1978,Coleman1975}.  
It is also worth mentioning as a technical point that we can now see why  nearest-neighbor interaction alone won't give rise to the SSH  phase,  which requires $g<0$ and $K<1/2$.  

\subsection{Edge states (or lack thereof)}
\label{single_topological}
Let us now turn to the topological properties of the model \eqref{Hsingle},  which requires looking at a finite chain with open boundary conditions,  i.e.  edges.  As mentioned in the Introduction,  we are asking the question of how the degeneracy of the ground state is changed when the edges are added.  This generalizes the non-interacting notion of the edge states.  

In a previous work \cite{Matveeva2024},  we studied in detail how to identify edge states using bosonization.  As a summary,  the relevant boundary conditions (for our bosonization convention) are 
$\sqrt{4\pi} \phi = 0 \mod 2\pi$.   Edge states can then be examined by looking at semi-classical solitons between the boundary and bulk values of $\phi$.  Other works that focus on integrable models have verified the validity of this approach \cite{Andrei2020,Andrei2025}.

Let us start with a charge-density wave.  Possible ground state field configurations are shown in Fig.  \ref{single_chain_phases_a}.  
It is important to remember that there is a gauge symmetry $\sqrt{4\pi} \phi \rightarrow \sqrt{4\pi} \phi  +2\pi$,  so the $y$ axes should be thought of as wrapped into a cylinder.  
The ground state is two-fold degenerate (the standard bulk $\mathbb{Z}_2$ spontaneous symmetry breaking) corresponding to the blue and red lines in Fig.  \ref{single_chain_phases_a}.   Each of these has a unique lowest-energy path to the boundary value of $\phi$.  In other words,  there is no extra degeneracy at the boundary; therefore,   there are no topological edge states.  

 The SSH phase also exhibits two degenerate states in the bulk that correspond to different bond dimerization patterns shifted by the lattice constant $a_0$ with respect to each other,  similar to the two dimerized phases of the well-studied non-interacting SSH chain.  In the mean-field picture, one of these states (the blue one in the figure) is compatible with the boundary conditions and therefore is trivial.  However,  the other mean-field state (the red one in the figure) has two degenerate choices at each edge (the green lines) corresponding to edge solitons.  Therefore,  this mean-field state is topological and exhibits edge states.  This was analyzed in full detail in our previous paper \cite{Matveeva2024}.    One might therefore expect degenerate edge modes to also appear in the many-body bond-density wave phase.  However,  it turns out that this is not the case as we will now explain.  
 
\begin{figure}
\includegraphics[scale=0.13]{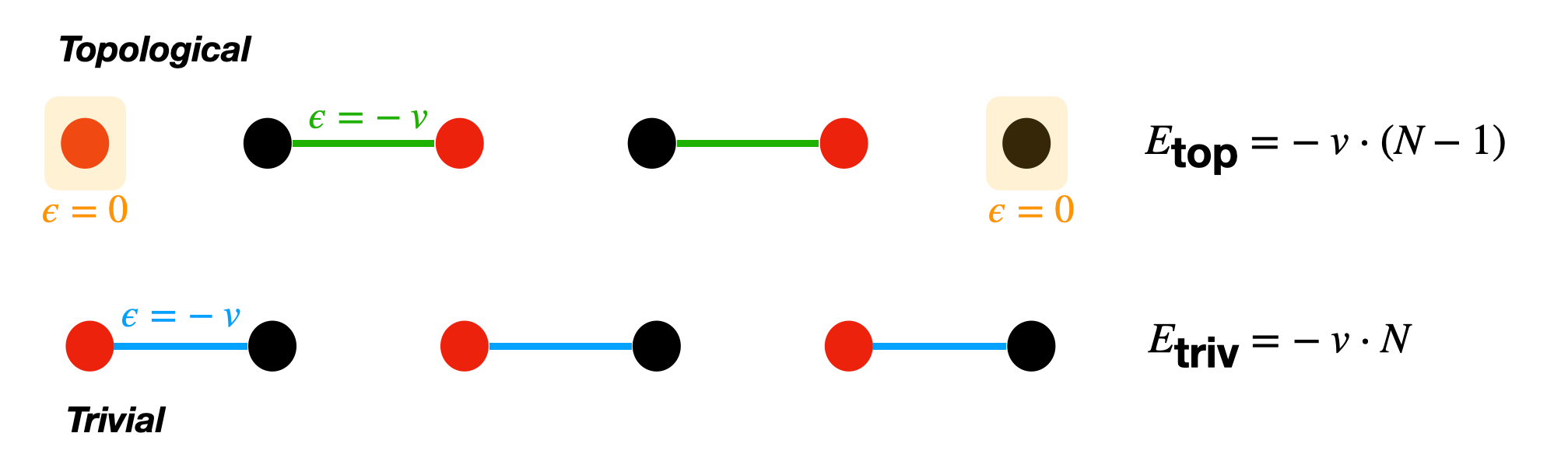}
\caption{Ground state energies of the two phases in a finite SSH model.  For a fixed number of particles $N$ at half filling,  the total ground state energy of the trivial phase is lower than the energy of a topological phase,  where one of the particles occupies one of the two zero-energy edge modes.  Here $\epsilon$ is the energy of a single electron localized on the bond between two atoms,  $v$ is the hopping amplitude.  For our interacting model (\ref{Hsingle}),  the energy scale $v$ is related to the interaction strength $g$.  }
\label{1chain_GS}
\end{figure}

Firstly,  looking carefully at Fig. \ref{single_chain_phases_b},  we see that the two ground states that were degenerate in the bulk are no longer degenerate in an open chain.  In particular,  the  blue state with no edge modes has lower energy than the red state,  as the green solitons cost energy.  There is a simple physical interpretation of this result,  which is illustrated in Fig. \ref{1chain_GS} for the extreme dimerized limit.  For a chain of length $2N$,  the ground state energy of the trivial dimerization pattern is $-v N$ while the presence of the edge states means that the ground state energy in the topological dimerization  pattern is $-v (N-1)$,  where $v$ is a parameter associated with the dimerization,  i.e. the gap.  In short,  the topological phase is higher in energy by $v$ than the trivial phase.  Looking back at Fig.  \ref{single_chain_phases_b},  we see that $v$ is the energy scale associated with the green solitons,  which shows that this energy difference between the bulk degenerate states remains beyond the extreme dimerization cartoon.   We emphasize that this is not the usual finite-size splitting between bulk degenerate symmetry-breaking states.  This is exponentially small in system size,  while this splitting is fixed energy and therefore scales as $1/L$ for the commonly used energy density.  
Secondly,  even if we force our system to be in the red bond-density wave state of Fig. \ref{single_chain_phases_b} (the phase in which the mean-field solution is topological),  there are no edge states.    This is because the green solitons at the edge in the figure can propagate into the bulk - these are the original fractionally charged solitons discussed in 1979 by Su,  Schrieffer,  and Heeger \cite{SSH1979,Kivelson1988}.   

In the mean-field model,  which is now commonly called the SSH model,  only one of the minima (red or blue) is the ground state in the bulk.  The other would be a local maximum.  However,  in the original model of SSH,  spontaneous symmetry breaking means that both states are local minima.   This is why the mean-field version has solitons confined to the edge,  while in the original model they propagate freely.  

It is worth noting that the CDW phase also has freely propagating solitons between the two different ground states, however,  these do not translate into edge states in the mean field version.  The underlying reason for this is that these states break chiral symmetry; however,  it is worthy of note that for modified boundary conditions, the CDW states also show edge states \cite{Azaria2025}.  It is unclear however,  how to realise such modified boundary conditions as they do not correspond to real edges of the chain.

\section{A spinful model}\label{Section_Model}
\subsection{The model}
From the simple spinless model studied in the previous section,  we now understand both the technical details and the new physics that may emerge from the interplay between conventional spontaneous symmetry breaking and topological edge states.  However, the end result was that none of the phases of this model had edge states.  Let us now turn to a spinful model,  where we show that this possibility may occur. 

We concentrate on models with spin-charge separation,  which is common in one dimension \cite{gogolin2004bosonization}.  The bosonized Hamiltonian is: 
\begin{align}
\label{full_model_main}
H=H_{\text{LL}} + V. 
\end{align}
Here $H_{\text{LL}}$ is the quadratic part of the Hamiltonian: 
\begin{align}
H_{\text{LL}} =\sum\limits_{\nu =c,s}  \frac{v_{\nu}}{2} \left[K_\nu \Pi^2_{\nu} +K^{-1}_{\nu} (\partial_x \phi_\nu)^2\right], 
\end{align}
where the fields $\phi_c$ and $\phi_s$ describe charge and spin degrees of freedom.  The gap opening $V$ term is given by: 
\begin{align}
\label{gap_terms_main}
V=\frac{g_c}{(\pi a)^2} \cos[\sqrt{8\pi} \phi_c]+\frac{g_s}{(\pi a)^2} \cos[\sqrt{8\pi} \phi_s],
\end{align}
where $a$ is the lattice constant.  In this work,  we concentrate on phases where both cosines are relevant,  which correspond to spontaneous symmetry breaking.   

For completeness,  a corresponding fermionic lattice model is discussed in detail in Appendix \ref{RG_analysis}.  As with the spinless case,  it is not completely trivial to write a model that realizes all relevant broken symmetry phases.   We believe that ours is one of the simplest,  as it describes spinful fermions with $U(1)$ spin symmetric interactions,  which could be useful for future numerical studies.  
It is worth noting that there has been a previous study \cite{Montorsi2017} of the topological phases of the model \eqref{full_model_main}.  Recent developments in understanding topological phases via bosonization make it prudent to re-examine these previous results. 

\subsection{Order parameters}\label{sec_phases}
\begin{figure}
\includegraphics[scale=0.13]{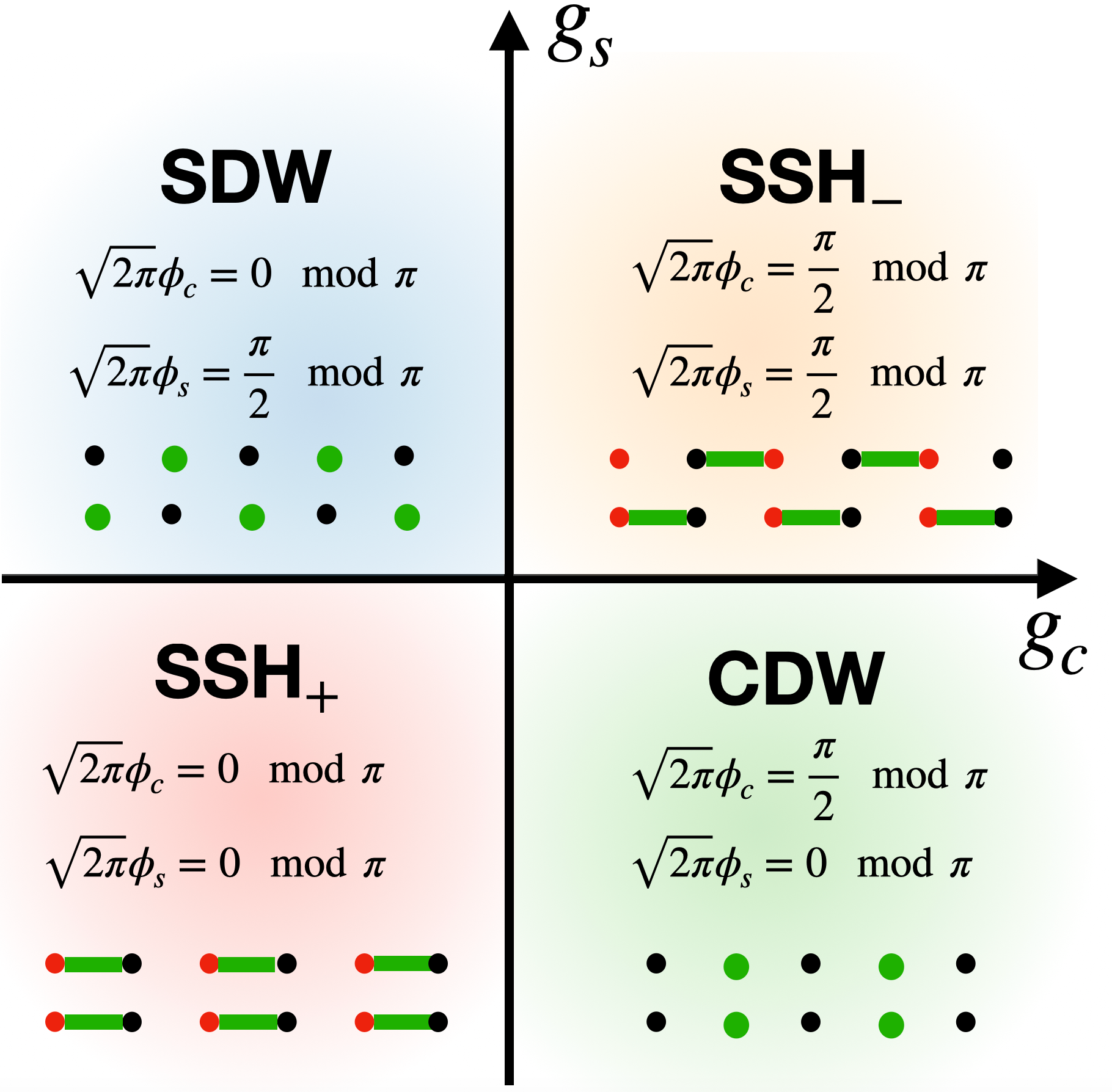}
\caption{Phase diagram of the model (\ref{full_model_main}).  Each gapped phase is characterized by one of the local order parameters given by (\ref{order_parameters_ferm}).  Green dots and lines schematically illustrate electronic densities in each of the phases. }
\label{phase_diagram}
\end{figure}

As before,  the phase diagram of the model \eqref{full_model_main} can be determined semi-classically by minimizing the gap opening part (\ref{gap_terms_main}) depending on the signs of $g_c$ and $g_s$.  The results are illustrated in Fig.  \ref{phase_diagram}.   We focus on the four fully gapped phases,  each of which is characterized by one of the local order parameters: 
\begin{align}
\label{order_parameters_ferm}
\mathcal{O}_{\text{CDW}}  &= (-1)^j \sum_{\sigma} c^{\dagger}_{j,\sigma} c_{j,\sigma} \propto
 \nonumber \\ &\propto \sin\left(\sqrt{2\pi} \phi_c\right) \cos\left(\sqrt{2\pi} \phi_s\right),  \nonumber \\
\mathcal{O}_{\text{SSH}_+} &= (-1)^j \sum_{\sigma} \left( c^{\dagger}_{j,\sigma} c_{j+1,\sigma} + \text{h.c.} \right) \propto \nonumber \\
 & \propto \cos\left(\sqrt{2\pi} \phi_c\right) \cos\left(\sqrt{2\pi} \phi_s\right),  \nonumber \\
\mathcal{O}_{\text{SSH}_-} &= (-1)^j \sum_{\sigma,\sigma'} \left( c^{\dagger}_{j,\sigma'} (\sigma_z)_{\sigma'\sigma} c_{j+1,\sigma} + \text{h.c.} \right)
  \propto \nonumber \\ 
  &\propto \sin\left(\sqrt{2\pi} \phi_c\right) \sin\left(\sqrt{2\pi} \phi_s\right),   \nonumber\\
\mathcal{O}_{\text{SDW}}  &= (-1)^j \sum_{\sigma,\sigma'} c^{\dagger}_{j,\sigma'} (\sigma_z)_{\sigma'\sigma} c_{j,\sigma}
  \propto \nonumber \\ &\propto \cos\left(\sqrt{2\pi} \phi_c\right) \sin\left(\sqrt{2\pi} \phi_s\right). 
\end{align}
The order parameters CDW and SDW characterize charge density and spin density wave phases,  correspondingly,  and the parameters $\text{SSH}_+$ and $\text{SSH}_-$  describe phases with bond dimerization (these are often called bond-density waves),  see Fig.  \ref{phase_diagram}.  

Each of the four phases is twofold degenerate (in an infinite system),   and each of the two ground states can be distinguished by different signs of a corresponding order parameter.  This is obvious from the strong-coupling illustrations in Fig.  \ref{phase_diagram}.  However, counting the states in bosonic language is a bit more subtle.  Following the previous discussion about a spinless chain,  one may think the degeneracy is four,  because there are two degenerate minima of the cosine in the charge sector  and two from the spin sector.  However,   simultaneous shift of $\sqrt{2\pi} \phi_c \rightarrow \sqrt{2\pi} \phi_c + \pi n $ and $\sqrt{2\pi} \phi_s \rightarrow \sqrt{2\pi} \phi_s + \pi k $,  where $n,k \in \mathbb{Z}$ does not change the fermionic operators and thus is gauge transformation.  This reduces the degeneracy to the expected two.  

\subsection{Chains with edges}
In order to study edge states,  we look at a finite chain with open boundary conditions,  i.e. edges.  The boundary conditions at the edge were derived in \cite{Matveeva2024},  where we obtained: 
\begin{align}
\label{impurity_fields_two}
& 1. \sqrt{2\pi} \phi_c =\sqrt{2\pi} \phi_s= 0 \mod 2\pi,   \hspace{0.2cm} \text{or}  \nonumber \\
& 2.  \sqrt{2\pi} \phi_c = \sqrt{2\pi} \phi_s=\pi \mod 2\pi.
\end{align}
As with the bulk states,  one needs to take into account the gauge symmetry that connects the charge and spin sectors in order not to over-count the edge modes.   
 
\subsubsection{Charge-density wave phase $g_c>0,g_s<0$}
\begin{figure*}[t]
\begin{subfigure}[b]{0.45\linewidth}
\includegraphics[scale=0.11]{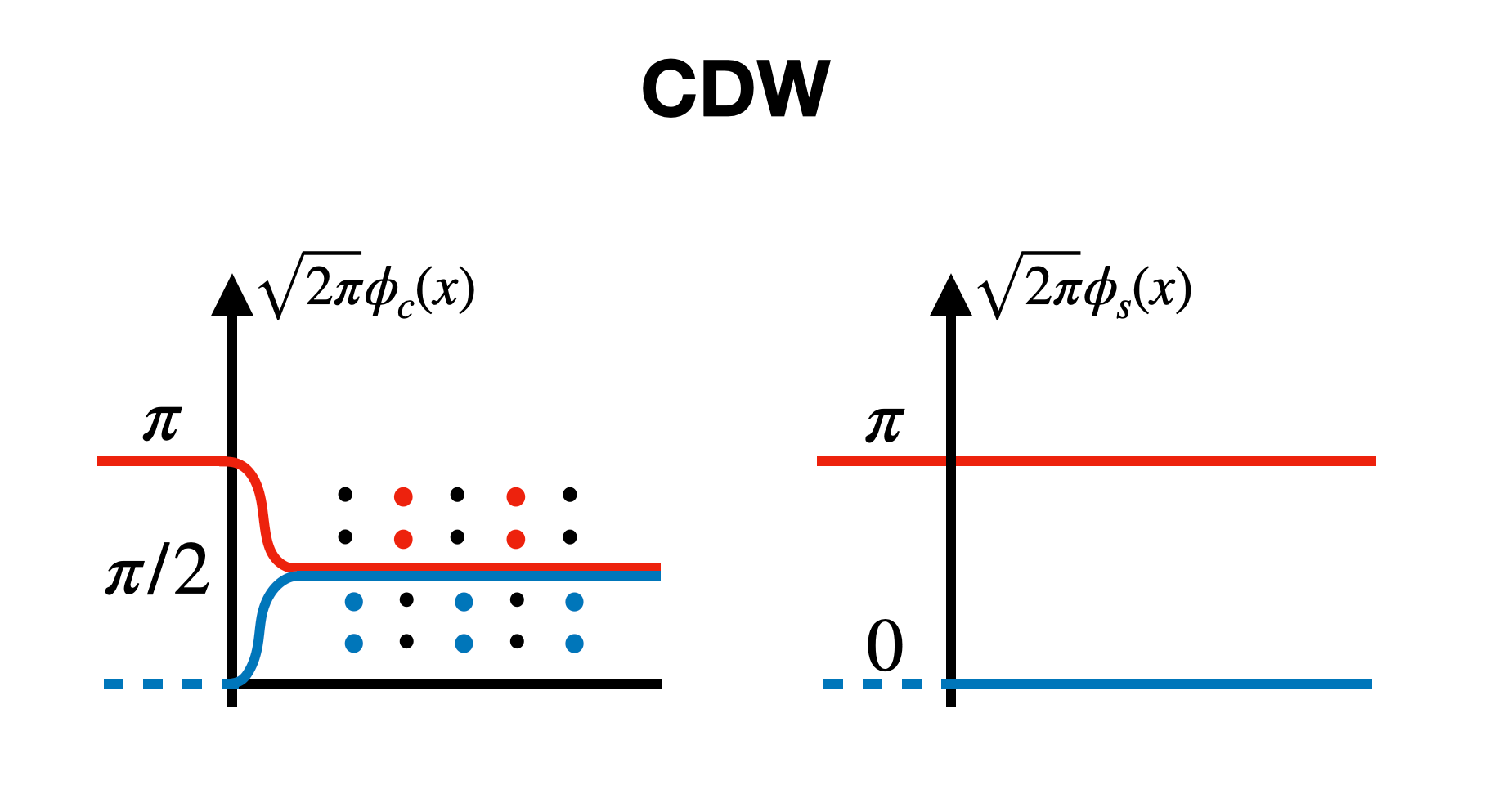}
\caption{}
\label{CDW_pic}
\end{subfigure}
\begin{subfigure}[b]{0.45\linewidth}
\includegraphics[scale=0.11]{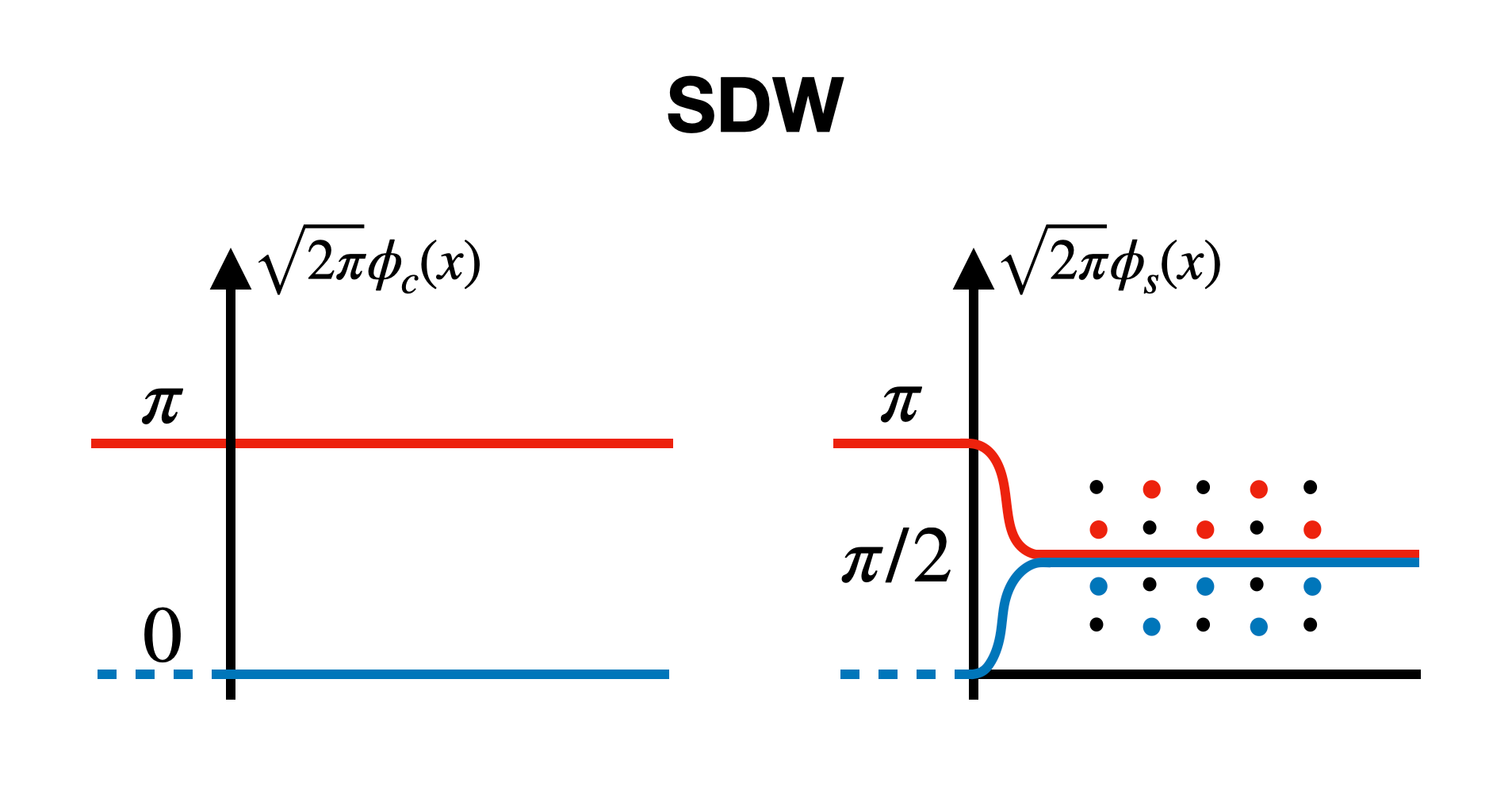}
\caption{}
\label{SDW_pic}
\end{subfigure}
\vskip\baselineskip
\begin{subfigure}[b]{0.45\linewidth}
\includegraphics[scale=0.11]{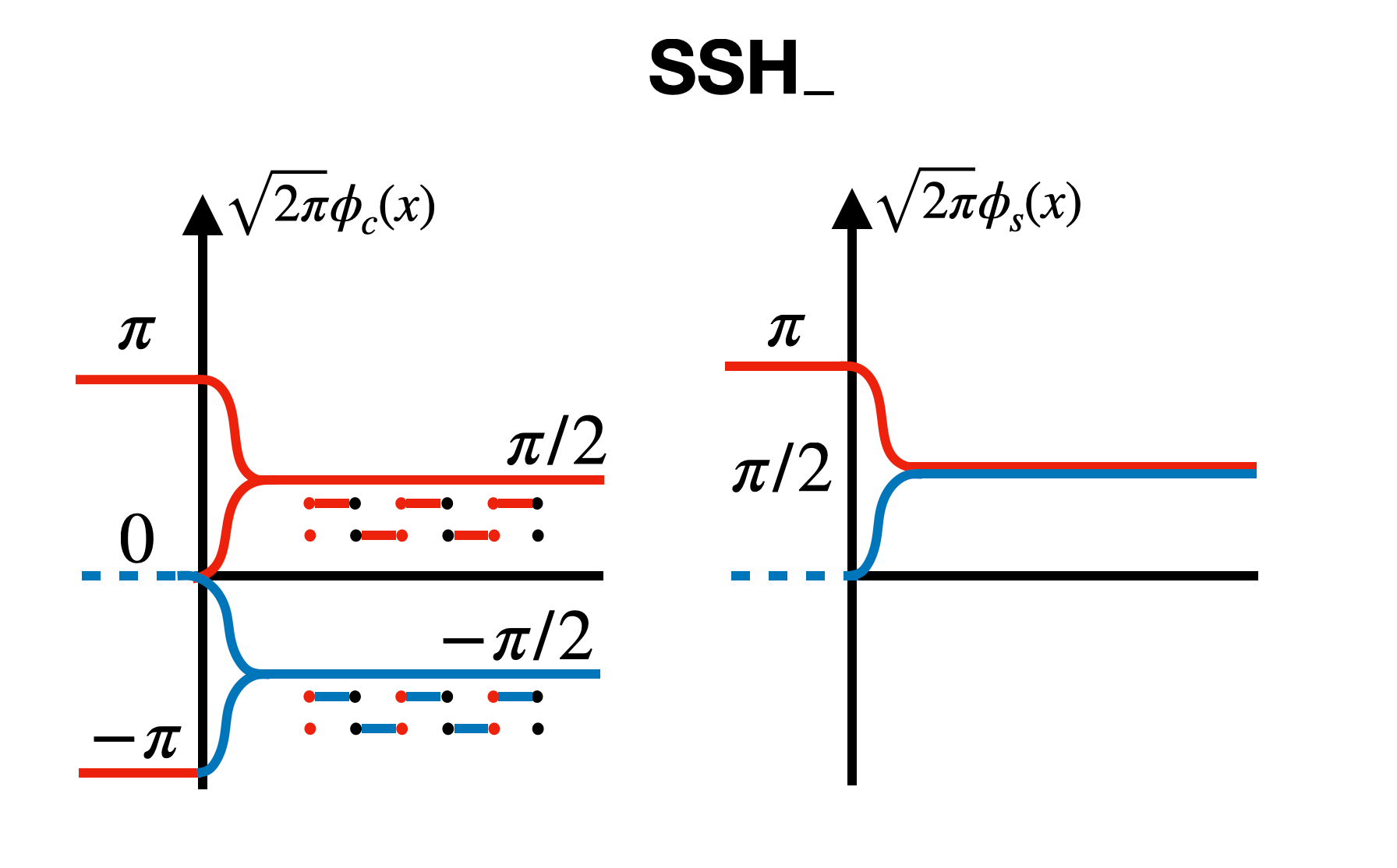}
\caption{}
\label{SSHm_pic}
\end{subfigure}
\begin{subfigure}[b]{0.45\linewidth}
\includegraphics[scale=0.11]{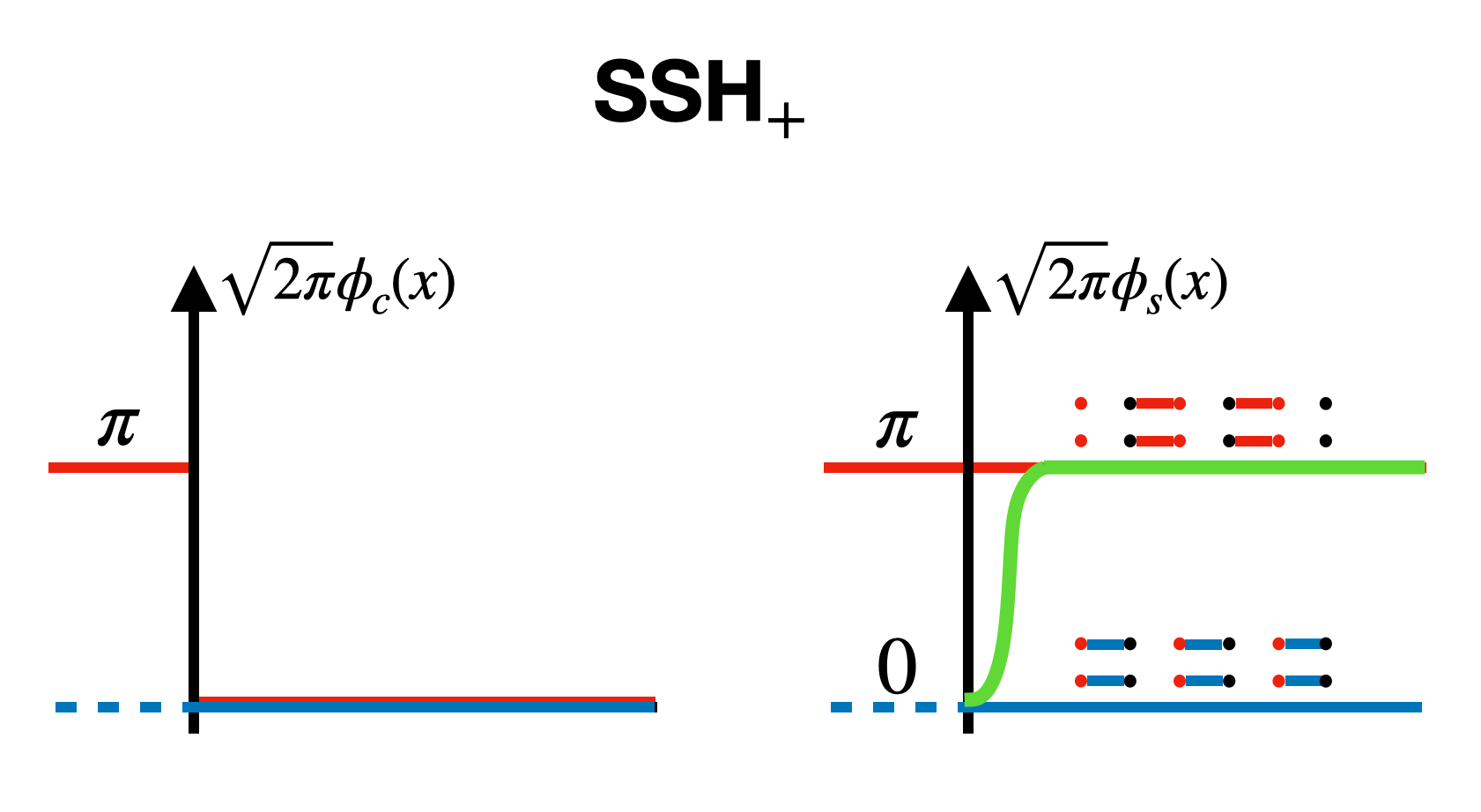}
\caption{}
\label{SSHp_pic}
\end{subfigure}
\caption{Profiles of the charge and spin fields in different gapped phases of the bosonic model (\ref{gap_terms_main}).  The ground state configuration is characterized by the bosonic kinks with the smallest magnitudes,  shown in red and blue solid lines.   }
\label{gapped_phases_fig}
\end{figure*}

In the charge density wave phase,  the physically distinct values of the bosonic fields in the ground state can be chosen as $\left(\sqrt{2\pi} \phi_c, \sqrt{2\pi} \phi_s \right)= (\pi/2, 0)$,  the blue line in the Fig.  \ref{CDW_pic} and $\left(\sqrt{2\pi} \phi_c, \sqrt{2\pi} \phi_s \right)= (\pi/2, \pi)$ is the red line in Fig.   \ref{CDW_pic}.   In the former case,  the minimum energy configuration goes to $(\pi, \pi)$ at the boundary,  while in the latter it goes to $(0,0)$.  Any other configuration will have higher energy and thus is not a ground state. 

Therefore,  two bulk states remain degenerate in the presence of boundaries (up to the usual exponentially small quantum splitting - this goes beyond semi-classical analysis),  and there is no extra degeneracy associated with the edge.  Therefore,  this phase is topologically trivial.

\subsubsection{Spin-density wave phase $g_c<0,g_s>0$}
In the spin density wave phase the ground state is given by $\left(\sqrt{2\pi} \phi_c, \sqrt{2\pi} \phi_s \right)= (0, \pi/2)$ and $\left(\sqrt{2\pi} \phi_c, \sqrt{2\pi} \phi_s \right)= (\pi/2, \pi)$.   Again,  similar to the charge density wave case,  there is no extra degeneracy related to topology,  as illustrated in Fig.  \ref{SDW_pic}.

\subsubsection{$\text{SSH}_-$ phase $g_c>0,g_s>0$}
The bulk bosonic ground state in the case when both coupling constants are positive: $g_c>0$ and $g_s>0$ can be chosen as: $\left(\sqrt{2\pi} \phi_c, \sqrt{2\pi} \phi_s \right)= (\pi/2, \pi/2)$ and $\left(\sqrt{2\pi} \phi_c, \sqrt{2\pi} \phi_s \right)= (-\pi/2, \pi/2)$. 
The ground state in an open system is shown in Fig.  \ref{SSHm_pic}.  Connecting those states to the edge values $(0,0)$ or $(\pi,\pi)$ requires a half-soliton in the charge and spin sector and is therefore degenerate.  

The total degeneracy in the presence of a single edge is therefore four,  higher than the two in the bulk.  The extra factor of two is associated with the presence of an edge mode and is illustrated in Fig.  \ref{SSHm_pic} by two red and two blue edge solitons in the charge sector and the corresponding solitons in the spin sector.   

This can be understood by thinking about the mean-field solution.  Unlike in the spinless SSH chain that has a trivial and a topological phase,   two mean-field solutions in the $\text{SSH}_-$ phase are both topological with winding number $\nu=1$. 

\subsubsection{$\text{SSH}_+$ phase $g_c<0,g_s<0$}
In the $\text{SSH}_-$ phase the ground state is given by $\left(\sqrt{2\pi} \phi_c, \sqrt{2\pi} \phi_s \right)= (0, 0)$ and $\left(\sqrt{2\pi} \phi_c, \sqrt{2\pi} \phi_s \right)= (0, \pi)$ and illustrated in Fig.   \ref{SSHp_pic}.  This phase is similar to the bond-density wave phase in the single-chain model discussed in Section \ref{single_section}, since the corresponding order parameter (\ref{order_parameters_ferm}) is the sum of two bond-density wave order parameters for each spin projection.  The ground state of an open chain is not degenerate, which can be explained by the same physical arguments presented in Section \ref{single_section}.

The lowest excited state is represented by a green soliton in Fig.   \ref{SSHp_pic},  which interpolates between two degenerate bulk minima.  It carries spin $1/2$ and represents a dynamic excitation of the bosonic model,  which is not bound to the edge.

\section{Topological protection of $\text{SSH}_-$ phase}\label{protection_section}
Let us now prove that the $\text{SSH}_-$ phase falls into a category of SPT phases.  In particular,  we need to show that the edge states are robust with respect to small perturbations.   By doing that,  we also need to identify the set of protective symmetries.  On the mean-field level,  this phase has a winding number $\nu=1$ and is protected by chiral symmetry.  We now show that it remains true beyond the mean-field theory.   

 In our previous works on weakly-interacting two-chain models  \cite{Matveeva2023,Matveeva2024},  we demonstrated that the chiral symmetry protects the degeneracy of  
the edge modes of the single-particle phases and for weakly-interacting models (where the interactions are smaller than the single-particle gap).  Let us now analyze its role in our strongly interacting case.  The chiral symmetry $S$ is one of three symmetries that fully determine the non-interacting topological classification.  It is given by the product of the time-reversal symmetry $T$ and the particle-hole symmetry $P$.  On the many-body level,  they are defined as follows \cite{Ludwig2016}: 

\begin{align}
\label{symmetries_many_body}
T^{-1} \hat{c}_{\alpha} T = (U^{\dagger}_T)_{\alpha,  \beta} \hat{c}_{\beta}\\
P^{-1} \hat{c}_{\alpha} P = (U^{*\dagger}_P)_{\alpha,  \beta} \hat{c}^{\dagger}_{\beta}\\
S^{-1} \hat{c}_{\alpha} S = (U^{*\dagger}_S)_{\alpha,  \beta}\hat{c}^{\dagger}_{\beta},  
\end{align}
where $U_S= U_T U_P^*$.  In the case of a spinful lattice model,  degrees of freedom $\alpha$ include both sublattice degrees of freedom and spin degrees of freedom.   Sublattices $A$ and $B$ are chosen such that $c_{j} \equiv c_{A,n}$ if $j$ is an even site and $c_{j} \equiv c_{B,n}$ if $j$ is an odd site.   Here $n$ is the index of a unit cell.   In addition,  in the many-body representation,  the time-reversal symmetry is an anti-unitary operator,  and the particle-hole is a unitary one.  

For the spinful model, the operators are represented by the following unitary matrices: 

\begin{equation}
\label{time-reversal}
\begin{cases}
 T^2=+1 : U^{+}_T=S_0 \sigma_x \\
 T^2=-1 : U^{-}_T=iS_0 \sigma_y,
\end{cases} 
\end{equation}
here Pauli matrices $S_i$ describe sublattice degrees of freedom,  and $\sigma_i$ correspond to spin degrees of freedom.  The particle-hole symmetry is represented by the following matrices: 
\begin{equation}
\label{particle-hole}
\begin{cases}
 P^2=+1 : U^{+}_P=S_z \sigma_x \\
P^2=-1 : U^{-}_P=-iS_z \sigma_y. 
\end{cases}
\end{equation}
Thus,  for chiral symmetry,  one has two options: 
\begin{equation}
\label{symmetries_representation}
\begin{cases}
U_{S_1} =  S_z\sigma_0,  \\
U_{S_2} = S_z\sigma_z.
\end{cases}
\end{equation}
As we discussed in \cite{Matveeva2023},  the chiral symmetry $S_1$ protects classes with integer classification,  while $S_2$ is needed to protect the $\mathbb{Z}_2$ topological phase in the class with time-reversal symmetry.    Since we are using bosonization to study topological properties,  we need to have the bosonic representation of the symmetries.  We derived them in \cite{Matveeva2024} and the results are summarized in Table \ref{charge_spin_symm_main}.  

\begin{table}
\begin{tabular}{c|cccccc}
\toprule
\multicolumn{1}{c}{} & \multicolumn{6}{c}{\textbf{Symmetry action $ \tilde{\phi}_j \equiv S^{-1} \phi_j S$}}  \\
\cmidrule(rl){2-7} 
\textbf{ } & {$T_+$} & {$T_-$} & {$P_+$} & {$P_-$} & {$S_1$}& {$S_2$}  \\
\hline
$ \tilde{\phi}_c$ & $\phi_c$& ($\phi_c +\sqrt{\frac{\pi}{2}}$) & $-\phi_c$ &$- (\phi_c + \sqrt{\frac{\pi}{2}})$ & $-\phi_c$ & $-(\phi_c +\sqrt{\frac{\pi}{2}})$ \\
\hline
$ \tilde{\phi}_s$ & $-\phi_s$ & $-(\phi_s+\sqrt{\frac{\pi}{2}})$ &  $\phi_s$& $(-\phi_s +\sqrt{\frac{\pi}{2}})$ & $-\phi_s$& $-(\phi_s+\sqrt{\frac{\pi}{2}})$\\
\bottomrule
\end{tabular}
\caption{Action of symmetry operators on bosonic fields in a spinful model.  $T_{\pm}$ denotes time-reversal symmetry:  $T^2_{\pm}=\pm 1$,   and $P_{\pm}$ denotes particle-hole symmetry: $P^2_{\pm}=\pm 1$.  Chiral symmetry is given by their product: $S_1=P_{\pm} T_{\pm}$,  $S_2=P_{\pm} T_{\mp}$.}
\label{charge_spin_symm_main}
\end{table}
A generic perturbation that breaks chiral symmetry $S_1$ is some linear combination of $ \cos(\sqrt{2\pi} n \phi_c)\sin(\sqrt{2\pi} m \phi_s)$ and  $ \cos(\sqrt{2\pi} n \phi_s)\sin(\sqrt{2\pi} m \phi_c)$,  where $n,m$ are integers.  One can check that a generic term of this form shifts the minima in the $\text{SSH}_-$ phase in one or two bosonic sectors and thus breaks the degeneracy of the edge modes.  At the same time,  any term that preserves chiral symmetry $S_1$ (but may break other symmetries) does not reduce the degeneracy of the edge states.  Such a term has the following general form: 
\begin{align}
V = \sum_{n,m=0}^\infty A_{n,m} \cos(\sqrt{2\pi} m \phi_c) \cos(\sqrt{2\pi} n \phi_s)
+\\  +B_{n,m}\sin(\sqrt{2\pi} n \phi_c) \sin(\sqrt{2\pi}m \phi_s)\,.
\end{align}
Our full model also preserves chiral symmetry $S_2$,  however adding a generic perturbation consistent with this symmetry,  for example,  $\sin(2\sqrt{2\pi}\phi_c)\cos(\sqrt{2\pi}\phi_s)$ or $\cos(\sqrt{2\pi}\phi_c)  \sin(2\sqrt{2\pi}\phi_s)$,  breaks the degeneracy of the edge modes.   
Therefore,  similar to the non-interacting case,  chiral symmetry $S_1$  is necessary to protect the degeneracy of the edge states.  
Note that the lattice interaction term (\ref{interactions}) proportional to $\sigma_0$  breaks chiral symmetry.   The symmetry can be restored if one shifts the density operators $n_{j, \sigma} \rightarrow  n_{j, \sigma} -1/2 $.  This shift,  however,  generates simple chemical potential terms that do not change the low-energy bosonic model and just shift the whole spectrum.

\section{Discussion and conclusion}
We studied topological properties of interacting one-dimensional models with spontaneously broken symmetries.  In these phases,  each bulk ground state corresponds to a ground state of a mean-field Hamiltonian with a certain topological index.  We showed that in a finite chain,  the energy density difference between the ground states corresponding to mean-field theories with different indices scales as $1/L$.    This is because the ground state energies of a trivial mean-field phase and a topological one differ by the energy of the edge modes (see Section \ref{single_topological}).  Furthermore,  if you force a system with a trivial ground state to the topological phase,  the mean-field edge states can propagate into the bulk as they are the solitons of the interacting model.  

We illustrated this first in a spinless interacting chain.  We showed that it features two phases that spontaneously break  $\mathbb{Z}_2$ symmetry: charge-density wave and bond-density wave (SSH phase).  In the bond-density wave phase,  the Hamiltonian can be deformed either to a trivial or topological non-interacting SSH chain.  We find that even though one of these mean-field theories is topologically non-trivial,  the many-body state is not degenerate and does not have edge modes.  
  
We then extended this analysis to a spinful model with $U(1)$ symmetric interactions.  We found four phases with spontaneously broken symmetry: charge-density wave,  spin-density wave,  and two bond-density wave states,  that we call $\text{SSH}_+$ and $\text{SSH}_-$.   In the  $\text{SSH}_+$ phase,  the model can be deformed to two identical SSH chains,  both trivial or topological.   In the  $\text{SSH}_-$ phase,  the model is equivalent to two chains in the opposite topological phases.   We found that similar to the spinless chain,  this $\text{SSH}_+$ state has no edge modes and thus is topologically trivial.   However,  since both mean-field theories for the $\text{SSH}_-$ phase are topologically non-trivial,  this many-body state is four-fold degenerate in the presence of a single edge.  We associate the extra  degeneracy, compared to the two-fold bulk degeneracy,  with a non-trivial topology of a model.   By using bosonic language,  we proved that the topology in this interacting topological state is protected by a chiral symmetry,  similar to the non-interacting case.  

We note that the topological properties of the same spinful model that we studied were previously analyzed in Ref.  \cite{Montorsi2017}.  While that work identifies the CDW and SDW phases as topological,  our approach leads to a different conclusion.
We argued that one needs to carefully take into account the gauge symmetry of the problem when computing the degeneracy of the edge states.  Because of this subtle point, we classify the CDW and SDW as topologically trivial.

In this work,  we focused exclusively on the fully gapped phases of the interacting model,  as these are the phases that exhibit spontaneous symmetry breaking.  However,  the transition lines between these phases are only partially gapped and support gapless modes.  There are a lot of works on gapless topological states in one dimension \cite{Keselman2015,Andrei2020,Kainaris2015,Kainaris2017,Verresen2021,Santos2016,Kainaris2018}.   It is a very interesting question whether these gapless topological states can be understood as a phase transition line between two gapped ones and how it depends on the topology of the gapped phases.  We will analyze this question in a forthcoming work \cite{gapless_future2025}.  

In one-dimensional models with two channels,  i.e.  spinful electrons analyzed here,  all fully gapped phases exhibit spontaneous symmetry breaking and therefore have a local order parameter and thus have a non-interacting mean-field picture.  With more than two channels,  this is no longer the case,  and one could construct non-trivial interacting phases that are not characterized by single-particle order parameters and thus cannot be deformed to non-interacting models.  For instance,  in three-chain models,  one may construct phases with non-trivial three-particle local order parameters,  as has been demonstrated in \cite{Santos2019}.  Such topological phases feature a non-trivial parafermionic fixed point with non-trivial statistics and a central charge that is different from the central charge of free particles.   
In general,  it has been shown that any gapless boundary between gapped phases in one dimension is described by a CFT \cite{Verresen2017}.  Using the approaches developed in this work,  one could construct microscopic chain models that feature such phases with non-trivial central charges.  This will be discussed elsewhere.  

We acknowledge useful discussions with Elio K\"onig.

\onecolumngrid
\appendix
\section*{Appendix} 
\section{Bosonisation conventions}\label{bosonization_conventions}
\label{bosonization_spinful}
In the low energy limit,  the fermionic operator for a given spin index is decomposed into right- and left-moving parts as follows: 
\begin{align}
c_{j,\sigma} \rightarrow \Psi_{\sigma}(x) = e^{ik_Fx} R_{\sigma}(x) + e^{-ik_Fx} L_{\sigma}(x)
\end{align}
The mapping between the chiral fermionic operators and bosonic fields is given by: 
\begin{align}
\label{RL_bosonizedS}
R_{\sigma}(x) = \frac{\kappa_{\sigma}}{\sqrt{2\pi a}} e^{i\sqrt{4\pi} \phi_{R\sigma}},  \hspace{0.3cm} L_{\sigma}(x) = \frac{\kappa_{\sigma}}{\sqrt{2\pi a}} e^{-i\sqrt{4\pi} \phi_{L\sigma}}. 
\end{align}
The Klein factors $\kappa_{\sigma}$ are Hermitian operators and they satisfy the algebra $\{\kappa_{\sigma}, \kappa_{\sigma'}\}= 2\delta_{\sigma \sigma'}$.  They ensure the correct anticommutation relations between  fermionic operators with different spin indexes.  The specific representation satisfying this algebra can be chosen,  such as $\kappa_{\uparrow}\kappa_{\downarrow} = -\kappa_{\downarrow}\kappa_{\uparrow}=i$.   The commutation relations between the bosonic fields are chosen such that fields with different spin indices commute: 
 \begin{align}
\label{RLboson_commutators_spin}
[\phi_{R\sigma}(x), \phi_{L\sigma'}(y)] = \frac{i}{4} \delta_{\sigma,\sigma'}\nonumber \\ 
[\phi_{\eta, \sigma} (x), \phi_{\eta'\sigma'}(y)]=  \frac{i}{4} \eta \delta_{\eta, \eta'} \delta_{\sigma,\sigma'} \text{sign}(x-y).
\end{align}
It is useful to introduce the charge and spin bosonic fields: 
\begin{align}
\label{charge_spin_bosons}
\phi_c = \frac{\phi_{\uparrow}+\phi_{\downarrow}}{\sqrt{2}}, \hspace{0.3cm} \phi_s = \frac{\phi_{\uparrow}-\phi_{\downarrow}}{\sqrt{2}}.
\end{align}
One can use those fields to express the non-oscillatory part of charge and spin densities:  
\begin{align}
\rho_c(x) = \sum_{\sigma} \Psi^{\dagger}_{\sigma}(x) \Psi_{\sigma}(x)= \sqrt{\frac{2}{\pi}} \partial_x\phi_c, \nonumber\\ 
\rho_s (x)= \frac{1}{2} \sum_{\sigma\sigma'} \Psi_{\sigma}^{\dagger}(x) (\sigma_z)_{\sigma\sigma'}  \Psi_{\sigma'}(x)= \frac{1}{\sqrt{2\pi}} \partial_x\phi_s.  
\end{align}
\section{Microscopic fermionic spinful model}
\label{RG_analysis}
We focus on the following lattice model of interacting fermions: 
\begin{align}
\label{lattice_model}
H=H_0+H_{\text{int}},
\end{align}
where $H_0$ is a free part: 
\begin{align}
\label{free}
H_0 = -t \sum_{j,\sigma} \left(c^{\dagger}_{j,\sigma} c_{j+1, \sigma} +\text{h.c} \right),
\end{align}
where $\sigma= 1,2$ is the spin degree of freedom,  and $H_{\text{int}}$ is the interacting part: 
\begin{align}
\label{interactions}
H_{\text{int}}= \sum_{j,\alpha} g_{\alpha} \hat{c}_j^{\dagger} \sigma_\alpha \hat{c}_j  \cdot \hat{c}^{\dagger}_{j+1} \sigma_\alpha \hat{c}_{j+1} 
+ V_0 \sum_{j}  \hat{n}_{j} \sigma_0 \hat{n}_{j+1},
\end{align}
here Pauli matrices $\sigma_\alpha$ act on spin degrees of freedom and  $g_{z} \equiv g_{\parallel},  g_x=g_y \equiv g_{\perp}$,  $\sigma_0$ denotes an identity matrix.  We bosonize the full model at half filling $k_F=\pi/2$,  using rules summarized in the Appendix \ref{bosonization_spinful}.  We obtain the following bosonic model,  keeping the most relevant terms: 
\begin{align}
\label{full_model}
H=H_{\text{LL}} + V. 
\end{align}
Here $H_{\text{LL}}$ is the quadratic part of the Hamiltonian: 
\begin{align}
H_{\text{LL}} =\sum\limits_{\nu =c,s}  \frac{v_{\nu}}{2} \left[K_\nu \Pi^2_{\nu} +K^{-1}_{\nu} (\partial_x \phi_\nu)^2\right], 
\end{align}
where the fields $\phi_c$ and $\phi_s$ describe charge and spin degrees of freedom. 
The $V$ term in \eqref{full_model} is the gap opening part: 
\begin{align}
\label{gap_terms}
V=\frac{g_c}{(\pi a)^2} \cos[\sqrt{8\pi} \phi_c]+\frac{g_s}{(\pi a)^2} \cos[\sqrt{8\pi} \phi_s].
\end{align}
The relation between the microscopic parameters of the lattice model (\ref{lattice_model})  and the parameters of the low-energy model (\ref{full_model}) is the following: 
\begin{align}
\begin{cases}
K_c \simeq 1-g_{\parallel}/\pi v_F - 2g_{\perp}/\pi v_F - 2V_0/\pi v_F \\
K_s \simeq 1-3g_{\parallel}/\pi v_F + 2g_{\perp}/\pi v_F - 2V_0/\pi v_F.
\end{cases}
\end{align}
The coupling constants $g_s$ and $g_c$ in (\ref{gap_terms}) are given by: 
\begin{align}
g_s = g_{\parallel}-2g_{\perp} \\ 
g_c = -(g_{\parallel}+2g_{\perp}).
\end{align}
\begin{figure}
\includegraphics[scale=0.11]{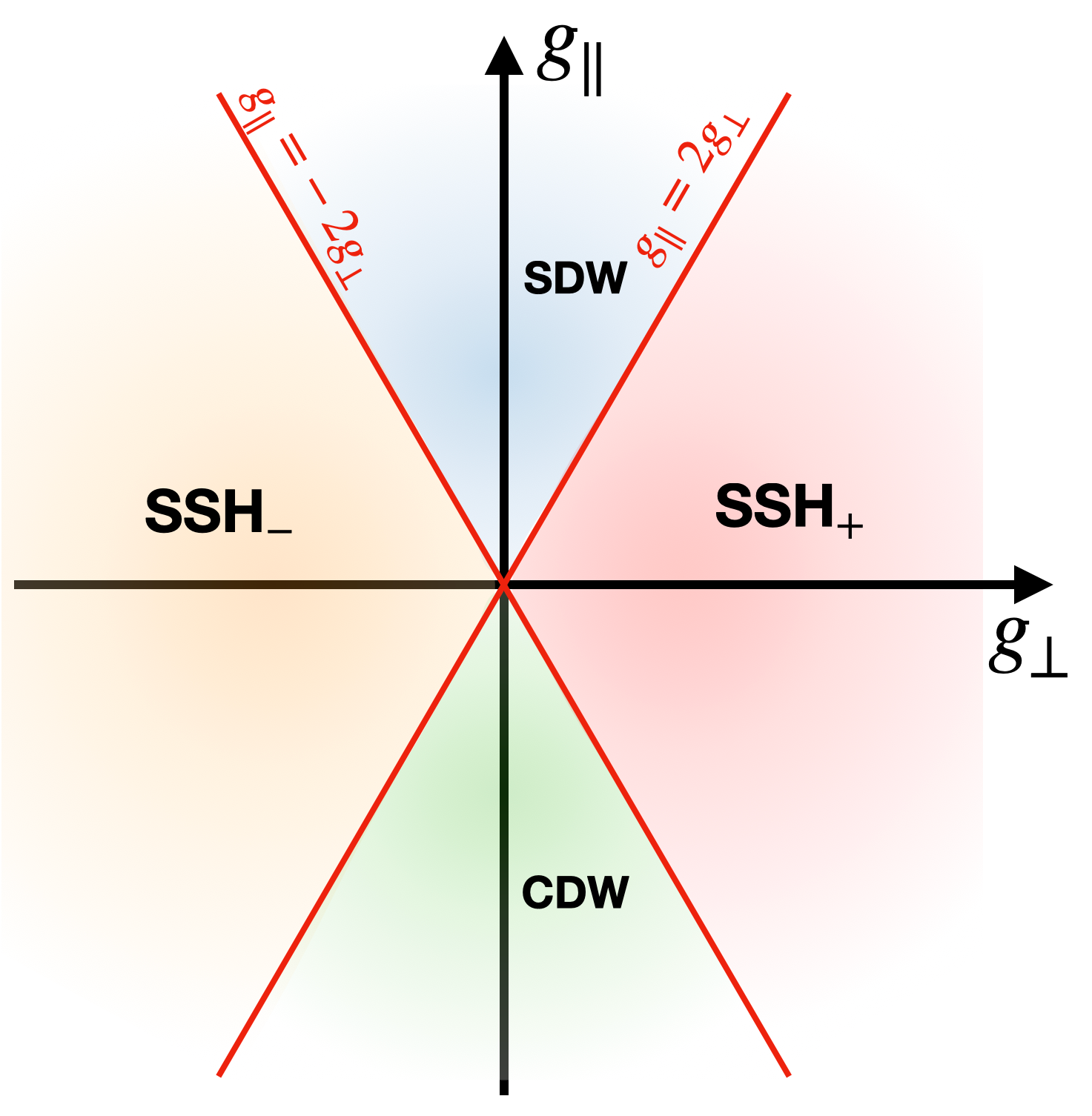}
\caption{Phase diagram of the model (\ref{full_model}) in terms of microscopic interaction parameters $g_{\parallel}$ and $g_{\perp}$.
 }
\label{phase_microscopic}
\end{figure}
Next,  we analyze the model (\ref{full_model}) using RG.  First,  we introduce the following dimensionless parameters: 
\begin{align}
\label{z_parameters}
\begin{cases}
z_{\perp,c}= -\frac{2}{\pi v_F} (g_{\parallel}+2g_{\perp})\\
z_{\parallel,c}=-\frac{2}{\pi v_F}(g_{\parallel}+2g_{\perp}+2V_0)
\end{cases}
\begin{cases}
z_{\perp,s}= \frac{2}{\pi v_F} (g_{\parallel}-2g_{\perp})\\
z_{\parallel,s}=-\frac{2}{\pi v_F}(3g_{\parallel}-2g_{\perp}+2V_0).
\end{cases}
\end{align}
In terms of these variables,  the RG equations take the following BKT-like form \cite{gogolin2004bosonization}:
\begin{align}
\label{RG_equations}
\frac{dz_{\parallel, \nu}}{dl} = - z^2_{\perp,  \nu},  
\frac{dz_{\perp,  \nu}}{dl} = - z_{\perp,\nu} \cdot z_{\parallel,\nu}. 
\end{align} 
Note that the parameters (\ref{z_parameters}) are not completely independent.   It is convenient to express  them as follows: 
\begin{align}
\label{z_parameters_indep}
\begin{cases}
z_{\parallel,c}=z_{\perp,c}-4V_0/\pi v_F \\ 
z_{\parallel,s}=-2z_{\perp,s}+z_{\parallel,c},
\end{cases}
\end{align}
where $z_{\perp,c}$ and $z_{\perp,s}$ can be thought as independent.   
By changing $V_0$,  we may always tune $z_{\parallel,c}$ and $z_{\parallel,s}$ to be negative,  so the gap-opening terms are relevant.  The phase diagram in terms of the original microscopic parameters $g_{\parallel}$ and $g_{\perp}$ is shown in Fig.  \ref{phase_microscopic}.  In the Section \ref{sec_phases} of the main text,  we discussed these four gapped phases that correspond to bond-density,  spin-density and charge-density wave orders.  Let us consider here simple limits of the microscopic model (\ref{lattice_model}) that would help us understand these phases more intuitively.  

\begin{itemize}
\item To understand the spin-density wave and charge density wave  phases,  we may set $g_{\perp}=0$ and $g_{\parallel} \neq 0$.   In that case,  the spin-dependent part of the interacting term (\ref{interactions}) is just proportional to the Ising term $S_{z,j}\cdot S_{z,j+1}$ that for $g_{\parallel}>0$ describes antiferromagnetic coupling.  That is precisely the spin-density wave type of order.  In the ferromagnetic case,  when $g_{\parallel}<0$,  we also need to take into account the density-density interaction term,   as without that term the model would be gapless,  according to RG equations (\ref{RG_equations}).  In that case,  we need to focus on repulsive interactions $V_0>0$  and $V_0> |g_{\parallel}|$ to make our gap opening terms relevant.   Charge density wave order is energetically favourable in that limit,  which corresponds to strong nearest-neighbor interactions. 

\item In the limit $g_{\parallel}=0$ and $g_{\perp}\neq 0$ we obtain bond density wave phases.  In this regime the gap-opening interaction term is proportional to $\left(c^{\dagger}_{j,\uparrow}c_{j+1,\uparrow}\right)\cdot (c^{\dagger}_{j+1,\downarrow}c_{j,\downarrow}) +\text{h.c.}$.  On the mean field level this term favours bond-density wave states with $\langle c^{\dagger}_{j,\sigma}c_{j+1,\sigma}\rangle \neq 0$ that have the same or the opposite dimerization patterns for the two spin projections,  depending on the sign of $g_{\perp}$. 

\end{itemize}

\bibliography{Manuscript_v1}

\begin{thebibliography}{44}%
\makeatletter
\providecommand \@ifxundefined [1]{%
 \@ifx{#1\undefined}
}%
\providecommand \@ifnum [1]{%
 \ifnum #1\expandafter \@firstoftwo
 \else \expandafter \@secondoftwo
 \fi
}%
\providecommand \@ifx [1]{%
 \ifx #1\expandafter \@firstoftwo
 \else \expandafter \@secondoftwo
 \fi
}%
\providecommand \natexlab [1]{#1}%
\providecommand \enquote  [1]{``#1''}%
\providecommand \bibnamefont  [1]{#1}%
\providecommand \bibfnamefont [1]{#1}%
\providecommand \citenamefont [1]{#1}%
\providecommand \href@noop [0]{\@secondoftwo}%
\providecommand \href [0]{\begingroup \@sanitize@url \@href}%
\providecommand \@href[1]{\@@startlink{#1}\@@href}%
\providecommand \@@href[1]{\endgroup#1\@@endlink}%
\providecommand \@sanitize@url [0]{\catcode `\\12\catcode `\$12\catcode
  `\&12\catcode `\#12\catcode `\^12\catcode `\_12\catcode `\%12\relax}%
\providecommand \@@startlink[1]{}%
\providecommand \@@endlink[0]{}%
\providecommand \url  [0]{\begingroup\@sanitize@url \@url }%
\providecommand \@url [1]{\endgroup\@href {#1}{\urlprefix }}%
\providecommand \urlprefix  [0]{URL }%
\providecommand \Eprint [0]{\href }%
\providecommand \doibase [0]{https://doi.org/}%
\providecommand \selectlanguage [0]{\@gobble}%
\providecommand \bibinfo  [0]{\@secondoftwo}%
\providecommand \bibfield  [0]{\@secondoftwo}%
\providecommand \translation [1]{[#1]}%
\providecommand \BibitemOpen [0]{}%
\providecommand \bibitemStop [0]{}%
\providecommand \bibitemNoStop [0]{.\EOS\space}%
\providecommand \EOS [0]{\spacefactor3000\relax}%
\providecommand \BibitemShut  [1]{\csname bibitem#1\endcsname}%
\let\auto@bib@innerbib\@empty
\bibitem [{\citenamefont {Klitzing}\ \emph {et~al.}(1980)\citenamefont
  {Klitzing}, \citenamefont {Dorda},\ and\ \citenamefont
  {Pepper}}]{Klitzing1980}%
  \BibitemOpen
  \bibfield  {author} {\bibinfo {author} {\bibfnamefont {K.~v.}\ \bibnamefont
  {Klitzing}}, \bibinfo {author} {\bibfnamefont {G.}~\bibnamefont {Dorda}},\
  and\ \bibinfo {author} {\bibfnamefont {M.}~\bibnamefont {Pepper}},\
  }\bibfield  {title} {\bibinfo {title} {New method for high-accuracy
  determination of the fine-structure constant based on quantized {H}all
  resistance},\ }\href {https://doi.org/10.1103/PhysRevLett.45.494} {\bibfield
  {journal} {\bibinfo  {journal} {Phys. Rev. Lett.}\ }\textbf {\bibinfo
  {volume} {45}},\ \bibinfo {pages} {494} (\bibinfo {year} {1980})}\BibitemShut
  {NoStop}%
\bibitem [{\citenamefont {Thouless}\ \emph {et~al.}(1982)\citenamefont
  {Thouless}, \citenamefont {Kohmoto}, \citenamefont {Nightingale},\ and\
  \citenamefont {den Nijs}}]{Thouless1982}%
  \BibitemOpen
  \bibfield  {author} {\bibinfo {author} {\bibfnamefont {D.~J.}\ \bibnamefont
  {Thouless}}, \bibinfo {author} {\bibfnamefont {M.}~\bibnamefont {Kohmoto}},
  \bibinfo {author} {\bibfnamefont {M.~P.}\ \bibnamefont {Nightingale}},\ and\
  \bibinfo {author} {\bibfnamefont {M.}~\bibnamefont {den Nijs}},\ }\bibfield
  {title} {\bibinfo {title} {Quantized {H}all conductance in a two-dimensional
  periodic potential},\ }\href {https://doi.org/10.1103/PhysRevLett.49.405}
  {\bibfield  {journal} {\bibinfo  {journal} {Phys. Rev. Lett.}\ }\textbf
  {\bibinfo {volume} {49}},\ \bibinfo {pages} {405} (\bibinfo {year}
  {1982})}\BibitemShut {NoStop}%
\bibitem [{\citenamefont {Zirnbauer}(1996)}]{Zirnbauer1996}%
  \BibitemOpen
  \bibfield  {author} {\bibinfo {author} {\bibfnamefont {M.~R.}\ \bibnamefont
  {Zirnbauer}},\ }\bibfield  {title} {\bibinfo {title} {{Riemannian symmetric
  superspaces and their origin in random‐matrix theory}},\ }\href
  {https://doi.org/10.1063/1.531675} {\bibfield  {journal} {\bibinfo  {journal}
  {Journal of Mathematical Physics}\ }\textbf {\bibinfo {volume} {37}},\
  \bibinfo {pages} {4986} (\bibinfo {year} {1996})}\BibitemShut {NoStop}%
\bibitem [{\citenamefont {Altland}\ and\ \citenamefont
  {Zirnbauer}(1997)}]{Altland1997}%
  \BibitemOpen
  \bibfield  {author} {\bibinfo {author} {\bibfnamefont {A.}~\bibnamefont
  {Altland}}\ and\ \bibinfo {author} {\bibfnamefont {M.~R.}\ \bibnamefont
  {Zirnbauer}},\ }\bibfield  {title} {\bibinfo {title} {Nonstandard symmetry
  classes in mesoscopic normal-superconducting hybrid structures},\ }\href
  {https://doi.org/10.1103/PhysRevB.55.1142} {\bibfield  {journal} {\bibinfo
  {journal} {Phys. Rev. B}\ }\textbf {\bibinfo {volume} {55}},\ \bibinfo
  {pages} {1142} (\bibinfo {year} {1997})}\BibitemShut {NoStop}%
\bibitem [{\citenamefont {Kitaev}(2009)}]{Kitaev2009}%
  \BibitemOpen
  \bibfield  {author} {\bibinfo {author} {\bibfnamefont {A.}~\bibnamefont
  {Kitaev}},\ }\bibfield  {title} {\bibinfo {title} {{Periodic table for
  topological insulators and superconductors}},\ }\href
  {https://doi.org/10.1063/1.3149495} {\bibfield  {journal} {\bibinfo
  {journal} {AIP Conference Proceedings}\ }\textbf {\bibinfo {volume} {1134}},\
  \bibinfo {pages} {22} (\bibinfo {year} {2009})}\BibitemShut {NoStop}%
\bibitem [{\citenamefont {Ryu}\ \emph {et~al.}(2010)\citenamefont {Ryu},
  \citenamefont {Schnyder}, \citenamefont {Furusaki},\ and\ \citenamefont
  {Ludwig}}]{Ryu2010}%
  \BibitemOpen
  \bibfield  {author} {\bibinfo {author} {\bibfnamefont {S.}~\bibnamefont
  {Ryu}}, \bibinfo {author} {\bibfnamefont {A.~P.}\ \bibnamefont {Schnyder}},
  \bibinfo {author} {\bibfnamefont {A.}~\bibnamefont {Furusaki}},\ and\
  \bibinfo {author} {\bibfnamefont {A.~W.~W.}\ \bibnamefont {Ludwig}},\
  }\bibfield  {title} {\bibinfo {title} {Topological insulators and
  superconductors: tenfold way and dimensional hierarchy},\ }\href
  {https://doi.org/10.1088/1367-2630/12/6/065010} {\bibfield  {journal}
  {\bibinfo  {journal} {New Journal of Physics}\ }\textbf {\bibinfo {volume}
  {12}},\ \bibinfo {pages} {065010} (\bibinfo {year} {2010})}\BibitemShut
  {NoStop}%
\bibitem [{\citenamefont {Morimoto}\ \emph {et~al.}(2015)\citenamefont
  {Morimoto}, \citenamefont {Furusaki},\ and\ \citenamefont
  {Mudry}}]{Morimoto2015}%
  \BibitemOpen
  \bibfield  {author} {\bibinfo {author} {\bibfnamefont {T.}~\bibnamefont
  {Morimoto}}, \bibinfo {author} {\bibfnamefont {A.}~\bibnamefont {Furusaki}},\
  and\ \bibinfo {author} {\bibfnamefont {C.}~\bibnamefont {Mudry}},\ }\bibfield
   {title} {\bibinfo {title} {Breakdown of the topological classification
  $\mathbb{Z}$ for gapped phases of noninteracting fermions by quartic
  interactions},\ }\href {https://doi.org/10.1103/PhysRevB.92.125104}
  {\bibfield  {journal} {\bibinfo  {journal} {Phys. Rev. B}\ }\textbf {\bibinfo
  {volume} {92}},\ \bibinfo {pages} {125104} (\bibinfo {year}
  {2015})}\BibitemShut {NoStop}%
\bibitem [{\citenamefont {Fidkowski}\ and\ \citenamefont
  {Kitaev}(2010)}]{Fidkowski2010}%
  \BibitemOpen
  \bibfield  {author} {\bibinfo {author} {\bibfnamefont {L.}~\bibnamefont
  {Fidkowski}}\ and\ \bibinfo {author} {\bibfnamefont {A.}~\bibnamefont
  {Kitaev}},\ }\bibfield  {title} {\bibinfo {title} {Effects of interactions on
  the topological classification of free fermion systems},\ }\href
  {https://doi.org/10.1103/PhysRevB.81.134509} {\bibfield  {journal} {\bibinfo
  {journal} {Phys. Rev. B}\ }\textbf {\bibinfo {volume} {81}},\ \bibinfo
  {pages} {134509} (\bibinfo {year} {2010})}\BibitemShut {NoStop}%
\bibitem [{\citenamefont {Fidkowski}\ and\ \citenamefont
  {Kitaev}(2011)}]{Fidkowski2011}%
  \BibitemOpen
  \bibfield  {author} {\bibinfo {author} {\bibfnamefont {L.}~\bibnamefont
  {Fidkowski}}\ and\ \bibinfo {author} {\bibfnamefont {A.}~\bibnamefont
  {Kitaev}},\ }\bibfield  {title} {\bibinfo {title} {Topological phases of
  fermions in one dimension},\ }\href
  {https://doi.org/10.1103/PhysRevB.83.075103} {\bibfield  {journal} {\bibinfo
  {journal} {Phys. Rev. B}\ }\textbf {\bibinfo {volume} {83}},\ \bibinfo
  {pages} {075103} (\bibinfo {year} {2011})}\BibitemShut {NoStop}%
\bibitem [{\citenamefont {Turner}\ \emph {et~al.}(2011)\citenamefont {Turner},
  \citenamefont {Pollmann},\ and\ \citenamefont {Berg}}]{Turner2011}%
  \BibitemOpen
  \bibfield  {author} {\bibinfo {author} {\bibfnamefont {A.~M.}\ \bibnamefont
  {Turner}}, \bibinfo {author} {\bibfnamefont {F.}~\bibnamefont {Pollmann}},\
  and\ \bibinfo {author} {\bibfnamefont {E.}~\bibnamefont {Berg}},\ }\bibfield
  {title} {\bibinfo {title} {Topological phases of one-dimensional fermions: An
  entanglement point of view},\ }\href
  {https://doi.org/10.1103/PhysRevB.83.075102} {\bibfield  {journal} {\bibinfo
  {journal} {Phys. Rev. B}\ }\textbf {\bibinfo {volume} {83}},\ \bibinfo
  {pages} {075102} (\bibinfo {year} {2011})}\BibitemShut {NoStop}%
\bibitem [{\citenamefont {Gu}\ and\ \citenamefont {Wen}(2009)}]{Wen2009}%
  \BibitemOpen
  \bibfield  {author} {\bibinfo {author} {\bibfnamefont {Z.-C.}\ \bibnamefont
  {Gu}}\ and\ \bibinfo {author} {\bibfnamefont {X.-G.}\ \bibnamefont {Wen}},\
  }\bibfield  {title} {\bibinfo {title} {Tensor-entanglement-filtering
  renormalization approach and symmetry-protected topological order},\ }\href
  {https://doi.org/10.1103/PhysRevB.80.155131} {\bibfield  {journal} {\bibinfo
  {journal} {Phys. Rev. B}\ }\textbf {\bibinfo {volume} {80}},\ \bibinfo
  {pages} {155131} (\bibinfo {year} {2009})}\BibitemShut {NoStop}%
\bibitem [{\citenamefont {Su}\ \emph {et~al.}(1979)\citenamefont {Su},
  \citenamefont {Schrieffer},\ and\ \citenamefont {Heeger}}]{SSH1979}%
  \BibitemOpen
  \bibfield  {author} {\bibinfo {author} {\bibfnamefont {W.~P.}\ \bibnamefont
  {Su}}, \bibinfo {author} {\bibfnamefont {J.~R.}\ \bibnamefont {Schrieffer}},\
  and\ \bibinfo {author} {\bibfnamefont {A.~J.}\ \bibnamefont {Heeger}},\
  }\bibfield  {title} {\bibinfo {title} {Solitons in polyacetylene},\ }\href
  {https://doi.org/10.1103/PhysRevLett.42.1698} {\bibfield  {journal} {\bibinfo
   {journal} {Phys. Rev. Lett.}\ }\textbf {\bibinfo {volume} {42}},\ \bibinfo
  {pages} {1698} (\bibinfo {year} {1979})}\BibitemShut {NoStop}%
\bibitem [{\citenamefont {Heeger}\ \emph {et~al.}(1988)\citenamefont {Heeger},
  \citenamefont {Kivelson}, \citenamefont {Schrieffer},\ and\ \citenamefont
  {Su}}]{Kivelson1988}%
  \BibitemOpen
  \bibfield  {author} {\bibinfo {author} {\bibfnamefont {A.~J.}\ \bibnamefont
  {Heeger}}, \bibinfo {author} {\bibfnamefont {S.}~\bibnamefont {Kivelson}},
  \bibinfo {author} {\bibfnamefont {J.~R.}\ \bibnamefont {Schrieffer}},\ and\
  \bibinfo {author} {\bibfnamefont {W.~P.}\ \bibnamefont {Su}},\ }\bibfield
  {title} {\bibinfo {title} {Solitons in conducting polymers},\ }\href
  {https://doi.org/10.1103/RevModPhys.60.781} {\bibfield  {journal} {\bibinfo
  {journal} {Rev. Mod. Phys.}\ }\textbf {\bibinfo {volume} {60}},\ \bibinfo
  {pages} {781} (\bibinfo {year} {1988})}\BibitemShut {NoStop}%
\bibitem [{\citenamefont {Sirker}\ \emph {et~al.}(2014)\citenamefont {Sirker},
  \citenamefont {Maiti}, \citenamefont {Konstantinidis},\ and\ \citenamefont
  {Sedlmayr}}]{Sirker2014}%
  \BibitemOpen
  \bibfield  {author} {\bibinfo {author} {\bibfnamefont {J.}~\bibnamefont
  {Sirker}}, \bibinfo {author} {\bibfnamefont {M.}~\bibnamefont {Maiti}},
  \bibinfo {author} {\bibfnamefont {N.~P.}\ \bibnamefont {Konstantinidis}},\
  and\ \bibinfo {author} {\bibfnamefont {N.}~\bibnamefont {Sedlmayr}},\
  }\bibfield  {title} {\bibinfo {title} {Boundary fidelity and entanglement in
  the symmetry protected topological phase of the {SSH} model},\ }\href
  {https://doi.org/10.1088/1742-5468/2014/10/P10032} {\bibfield  {journal}
  {\bibinfo  {journal} {Journal of Statistical Mechanics: Theory and
  Experiment}\ }\textbf {\bibinfo {volume} {2014}},\ \bibinfo {pages} {P10032}
  (\bibinfo {year} {2014})}\BibitemShut {NoStop}%
\bibitem [{\citenamefont {Li}\ \emph {et~al.}(2014)\citenamefont {Li},
  \citenamefont {Xu},\ and\ \citenamefont {Chen}}]{Li2014}%
  \BibitemOpen
  \bibfield  {author} {\bibinfo {author} {\bibfnamefont {L.}~\bibnamefont
  {Li}}, \bibinfo {author} {\bibfnamefont {Z.}~\bibnamefont {Xu}},\ and\
  \bibinfo {author} {\bibfnamefont {S.}~\bibnamefont {Chen}},\ }\bibfield
  {title} {\bibinfo {title} {Topological phases of generalized
  {S}u-{S}chrieffer-{H}eeger models},\ }\href
  {https://doi.org/10.1103/PhysRevB.89.085111} {\bibfield  {journal} {\bibinfo
  {journal} {Phys. Rev. B}\ }\textbf {\bibinfo {volume} {89}},\ \bibinfo
  {pages} {085111} (\bibinfo {year} {2014})}\BibitemShut {NoStop}%
\bibitem [{\citenamefont {Padavi\ifmmode~\acute{c}\else \'{c}\fi{}}\ \emph
  {et~al.}(2018)\citenamefont {Padavi\ifmmode~\acute{c}\else \'{c}\fi{}},
  \citenamefont {Hegde}, \citenamefont {DeGottardi},\ and\ \citenamefont
  {Vishveshwara}}]{DeGottardi2018}%
  \BibitemOpen
  \bibfield  {author} {\bibinfo {author} {\bibfnamefont {K.}~\bibnamefont
  {Padavi\ifmmode~\acute{c}\else \'{c}\fi{}}}, \bibinfo {author} {\bibfnamefont
  {S.~S.}\ \bibnamefont {Hegde}}, \bibinfo {author} {\bibfnamefont
  {W.}~\bibnamefont {DeGottardi}},\ and\ \bibinfo {author} {\bibfnamefont
  {S.}~\bibnamefont {Vishveshwara}},\ }\bibfield  {title} {\bibinfo {title}
  {Topological phases, edge modes, and the hofstadter butterfly in coupled
  {S}u-{S}chrieffer-{H}eeger systems},\ }\href
  {https://doi.org/10.1103/PhysRevB.98.024205} {\bibfield  {journal} {\bibinfo
  {journal} {Phys. Rev. B}\ }\textbf {\bibinfo {volume} {98}},\ \bibinfo
  {pages} {024205} (\bibinfo {year} {2018})}\BibitemShut {NoStop}%
\bibitem [{\citenamefont {Yahyavi}\ \emph {et~al.}(2018)\citenamefont
  {Yahyavi}, \citenamefont {Saleem},\ and\ \citenamefont
  {Hetényi}}]{Yahyavi_2018}%
  \BibitemOpen
  \bibfield  {author} {\bibinfo {author} {\bibfnamefont {M.}~\bibnamefont
  {Yahyavi}}, \bibinfo {author} {\bibfnamefont {L.}~\bibnamefont {Saleem}},\
  and\ \bibinfo {author} {\bibfnamefont {B.}~\bibnamefont {Hetényi}},\
  }\bibfield  {title} {\bibinfo {title} {Variational study of the interacting,
  spinless {S}u–{S}chrieffer–{H}eeger model},\ }\href
  {https://doi.org/10.1088/1361-648X/aae0a4} {\bibfield  {journal} {\bibinfo
  {journal} {Journal of Physics: Condensed Matter}\ }\textbf {\bibinfo {volume}
  {30}},\ \bibinfo {pages} {445602} (\bibinfo {year} {2018})}\BibitemShut
  {NoStop}%
\bibitem [{\citenamefont {Zegarra}\ \emph {et~al.}(2019)\citenamefont
  {Zegarra}, \citenamefont {Candido}, \citenamefont {Egues},\ and\
  \citenamefont {Chen}}]{Zegarra2019}%
  \BibitemOpen
  \bibfield  {author} {\bibinfo {author} {\bibfnamefont {A.}~\bibnamefont
  {Zegarra}}, \bibinfo {author} {\bibfnamefont {D.~R.}\ \bibnamefont
  {Candido}}, \bibinfo {author} {\bibfnamefont {J.~C.}\ \bibnamefont {Egues}},\
  and\ \bibinfo {author} {\bibfnamefont {W.}~\bibnamefont {Chen}},\ }\bibfield
  {title} {\bibinfo {title} {Corroborating the bulk-edge correspondence in
  weakly interacting one-dimensional topological insulators},\ }\href
  {https://doi.org/10.1103/PhysRevB.100.075114} {\bibfield  {journal} {\bibinfo
   {journal} {Phys. Rev. B}\ }\textbf {\bibinfo {volume} {100}},\ \bibinfo
  {pages} {075114} (\bibinfo {year} {2019})}\BibitemShut {NoStop}%
\bibitem [{\citenamefont {Nersesyan}(2020)}]{Nersesyan2020}%
  \BibitemOpen
  \bibfield  {author} {\bibinfo {author} {\bibfnamefont {A.~A.}\ \bibnamefont
  {Nersesyan}},\ }\bibfield  {title} {\bibinfo {title} {Phase diagram of an
  interacting staggered {S}u-{S}chrieffer-{H}eeger two-chain ladder close to a
  quantum critical point},\ }\href
  {https://doi.org/10.1103/PhysRevB.102.045108} {\bibfield  {journal} {\bibinfo
   {journal} {Phys. Rev. B}\ }\textbf {\bibinfo {volume} {102}},\ \bibinfo
  {pages} {045108} (\bibinfo {year} {2020})}\BibitemShut {NoStop}%
\bibitem [{\citenamefont {Jin}\ \emph {et~al.}(2023)\citenamefont {Jin},
  \citenamefont {Ruggiero},\ and\ \citenamefont {Giamarchi}}]{Giamarchi2023}%
  \BibitemOpen
  \bibfield  {author} {\bibinfo {author} {\bibfnamefont {T.}~\bibnamefont
  {Jin}}, \bibinfo {author} {\bibfnamefont {P.}~\bibnamefont {Ruggiero}},\ and\
  \bibinfo {author} {\bibfnamefont {T.}~\bibnamefont {Giamarchi}},\ }\bibfield
  {title} {\bibinfo {title} {Bosonization of the interacting
  {S}u-{S}chrieffer-{H}eeger model},\ }\href
  {https://doi.org/10.1103/PhysRevB.107.L201111} {\bibfield  {journal}
  {\bibinfo  {journal} {Phys. Rev. B}\ }\textbf {\bibinfo {volume} {107}},\
  \bibinfo {pages} {L201111} (\bibinfo {year} {2023})}\BibitemShut {NoStop}%
\bibitem [{\citenamefont {Matveeva}\ \emph {et~al.}(2023)\citenamefont
  {Matveeva}, \citenamefont {Hewitt}, \citenamefont {Liu}, \citenamefont
  {Reddy}, \citenamefont {Gutman},\ and\ \citenamefont {Carr}}]{Matveeva2023}%
  \BibitemOpen
  \bibfield  {author} {\bibinfo {author} {\bibfnamefont {P.}~\bibnamefont
  {Matveeva}}, \bibinfo {author} {\bibfnamefont {T.}~\bibnamefont {Hewitt}},
  \bibinfo {author} {\bibfnamefont {D.}~\bibnamefont {Liu}}, \bibinfo {author}
  {\bibfnamefont {K.}~\bibnamefont {Reddy}}, \bibinfo {author} {\bibfnamefont
  {D.}~\bibnamefont {Gutman}},\ and\ \bibinfo {author} {\bibfnamefont {S.~T.}\
  \bibnamefont {Carr}},\ }\bibfield  {title} {\bibinfo {title} {One-dimensional
  noninteracting topological insulators with chiral symmetry},\ }\href
  {https://doi.org/10.1103/PhysRevB.107.075422} {\bibfield  {journal} {\bibinfo
   {journal} {Phys. Rev. B}\ }\textbf {\bibinfo {volume} {107}},\ \bibinfo
  {pages} {075422} (\bibinfo {year} {2023})}\BibitemShut {NoStop}%
\bibitem [{\citenamefont {Melo}\ \emph {et~al.}(2023)\citenamefont {Melo},
  \citenamefont {J\'unior}, \citenamefont {Chen}, \citenamefont {Mondaini},\
  and\ \citenamefont {Paiva}}]{Melo2023}%
  \BibitemOpen
  \bibfield  {author} {\bibinfo {author} {\bibfnamefont {P.~B.}\ \bibnamefont
  {Melo}}, \bibinfo {author} {\bibfnamefont {S.~a. A.~S.}\ \bibnamefont
  {J\'unior}}, \bibinfo {author} {\bibfnamefont {W.}~\bibnamefont {Chen}},
  \bibinfo {author} {\bibfnamefont {R.}~\bibnamefont {Mondaini}},\ and\
  \bibinfo {author} {\bibfnamefont {T.}~\bibnamefont {Paiva}},\ }\bibfield
  {title} {\bibinfo {title} {Topological marker approach to an interacting
  {S}u-{S}chrieffer-{H}eeger model},\ }\href
  {https://doi.org/10.1103/PhysRevB.108.195151} {\bibfield  {journal} {\bibinfo
   {journal} {Phys. Rev. B}\ }\textbf {\bibinfo {volume} {108}},\ \bibinfo
  {pages} {195151} (\bibinfo {year} {2023})}\BibitemShut {NoStop}%
\bibitem [{\citenamefont {Matveeva}\ \emph {et~al.}(2024)\citenamefont
  {Matveeva}, \citenamefont {Gutman},\ and\ \citenamefont
  {Carr}}]{Matveeva2024}%
  \BibitemOpen
  \bibfield  {author} {\bibinfo {author} {\bibfnamefont {P.}~\bibnamefont
  {Matveeva}}, \bibinfo {author} {\bibfnamefont {D.}~\bibnamefont {Gutman}},\
  and\ \bibinfo {author} {\bibfnamefont {S.~T.}\ \bibnamefont {Carr}},\
  }\bibfield  {title} {\bibinfo {title} {Weakly interacting one-dimensional
  topological insulators: A bosonization approach},\ }\href
  {https://doi.org/10.1103/PhysRevB.109.165436} {\bibfield  {journal} {\bibinfo
   {journal} {Phys. Rev. B}\ }\textbf {\bibinfo {volume} {109}},\ \bibinfo
  {pages} {165436} (\bibinfo {year} {2024})}\BibitemShut {NoStop}%
\bibitem [{\citenamefont {Kitaev}(2001)}]{Kitaev2001}%
  \BibitemOpen
  \bibfield  {author} {\bibinfo {author} {\bibfnamefont {A.~Y.}\ \bibnamefont
  {Kitaev}},\ }\bibfield  {title} {\bibinfo {title} {Unpaired {M}ajorana
  fermions in quantum wires},\ }\href
  {https://doi.org/10.1070/1063-7869/44/10S/S29} {\bibfield  {journal}
  {\bibinfo  {journal} {Physics-Uspekhi}\ }\textbf {\bibinfo {volume} {44}},\
  \bibinfo {pages} {131} (\bibinfo {year} {2001})}\BibitemShut {NoStop}%
\bibitem [{\citenamefont {{Microsoft Azure Quantum}}\ \emph
  {et~al.}(2025)\citenamefont {{Microsoft Azure Quantum}}, \citenamefont
  {Aghaee}, \citenamefont {Alcaraz~Ramirez} \emph {et~al.}}]{Microsoft2025}%
  \BibitemOpen
  \bibfield  {author} {\bibinfo {author} {\bibnamefont {{Microsoft Azure
  Quantum}}}, \bibinfo {author} {\bibfnamefont {M.}~\bibnamefont {Aghaee}},
  \bibinfo {author} {\bibfnamefont {A.}~\bibnamefont {Alcaraz~Ramirez}}, \emph
  {et~al.},\ }\bibfield  {title} {\bibinfo {title} {Interferometric single-shot
  parity measurement in $\mathrm{InAs}$--$\mathrm{Al}$ hybrid devices},\ }\href
  {https://www.nature.com/articles/s41586-024-08445-2} {\bibfield  {journal}
  {\bibinfo  {journal} {Nature}\ }\textbf {\bibinfo {volume} {638}},\ \bibinfo
  {pages} {651} (\bibinfo {year} {2025})}\BibitemShut {NoStop}%
\bibitem [{\citenamefont {Alicea}\ \emph {et~al.}(2011)\citenamefont {Alicea},
  \citenamefont {Oreg}, \citenamefont {Refael}, \citenamefont {von Oppen},\
  and\ \citenamefont {Fisher}}]{Alicea2011}%
  \BibitemOpen
  \bibfield  {author} {\bibinfo {author} {\bibfnamefont {J.}~\bibnamefont
  {Alicea}}, \bibinfo {author} {\bibfnamefont {Y.}~\bibnamefont {Oreg}},
  \bibinfo {author} {\bibfnamefont {G.}~\bibnamefont {Refael}}, \bibinfo
  {author} {\bibfnamefont {F.}~\bibnamefont {von Oppen}},\ and\ \bibinfo
  {author} {\bibfnamefont {M.~P.~A.}\ \bibnamefont {Fisher}},\ }\bibfield
  {title} {\bibinfo {title} {Non-{A}belian statistics and topological quantum
  information processing in 1{D} wire networks},\ }\href
  {https://www.nature.com/articles/nphys1915} {\bibfield  {journal} {\bibinfo
  {journal} {Nature Physics}\ }\textbf {\bibinfo {volume} {7}},\ \bibinfo
  {pages} {412} (\bibinfo {year} {2011})}\BibitemShut {NoStop}%
\bibitem [{\citenamefont {Ryu}\ and\ \citenamefont {Hatsugai}(2006)}]{Ryu2006}%
  \BibitemOpen
  \bibfield  {author} {\bibinfo {author} {\bibfnamefont {S.}~\bibnamefont
  {Ryu}}\ and\ \bibinfo {author} {\bibfnamefont {Y.}~\bibnamefont {Hatsugai}},\
  }\bibfield  {title} {\bibinfo {title} {Entanglement entropy and the berry
  phase in the solid state},\ }\href
  {https://doi.org/10.1103/PhysRevB.73.245115} {\bibfield  {journal} {\bibinfo
  {journal} {Phys. Rev. B}\ }\textbf {\bibinfo {volume} {73}},\ \bibinfo
  {pages} {245115} (\bibinfo {year} {2006})}\BibitemShut {NoStop}%
\bibitem [{\citenamefont {Gogolin}\ \emph {et~al.}(2004)\citenamefont
  {Gogolin}, \citenamefont {Nersesyan},\ and\ \citenamefont
  {Tsvelik}}]{gogolin2004bosonization}%
  \BibitemOpen
  \bibfield  {author} {\bibinfo {author} {\bibfnamefont {A.}~\bibnamefont
  {Gogolin}}, \bibinfo {author} {\bibfnamefont {A.}~\bibnamefont {Nersesyan}},\
  and\ \bibinfo {author} {\bibfnamefont {A.}~\bibnamefont {Tsvelik}},\ }\href
  {https://books.google.co.il/books?id=BZDfFIpCoaAC} {\emph {\bibinfo {title}
  {Bosonization and Strongly Correlated Systems}}}\ (\bibinfo  {publisher}
  {Cambridge University Press},\ \bibinfo {year} {2004})\BibitemShut {NoStop}%
\bibitem [{\citenamefont {Zamolodchikov}\ and\ \citenamefont
  {Zamolodchikov}(1978)}]{Zamolodchikov1978}%
  \BibitemOpen
  \bibfield  {author} {\bibinfo {author} {\bibfnamefont {A.~B.}\ \bibnamefont
  {Zamolodchikov}}\ and\ \bibinfo {author} {\bibfnamefont {A.~B.}\ \bibnamefont
  {Zamolodchikov}},\ }\bibfield  {title} {\bibinfo {title} {Relativistic
  factorized s-matrix in two dimensions having o(n) isotopic symmetry},\ }\href
  {https://doi.org/https://doi.org/10.1016/0550-3213(78)90239-0} {\bibfield
  {journal} {\bibinfo  {journal} {Nuclear Physics B}\ }\textbf {\bibinfo
  {volume} {133}},\ \bibinfo {pages} {525} (\bibinfo {year}
  {1978})}\BibitemShut {NoStop}%
\bibitem [{\citenamefont {Coleman}(1975)}]{Coleman1975}%
  \BibitemOpen
  \bibfield  {author} {\bibinfo {author} {\bibfnamefont {S.}~\bibnamefont
  {Coleman}},\ }\bibfield  {title} {\bibinfo {title} {Quantum sine-{G}ordon
  equation as the massive thirring model},\ }\href
  {https://doi.org/10.1103/PhysRevD.11.2088} {\bibfield  {journal} {\bibinfo
  {journal} {Phys. Rev. D}\ }\textbf {\bibinfo {volume} {11}},\ \bibinfo
  {pages} {2088} (\bibinfo {year} {1975})}\BibitemShut {NoStop}%
\bibitem [{\citenamefont {Pasnoori}\ \emph {et~al.}(2020)\citenamefont
  {Pasnoori}, \citenamefont {Andrei},\ and\ \citenamefont
  {Azaria}}]{Andrei2020}%
  \BibitemOpen
  \bibfield  {author} {\bibinfo {author} {\bibfnamefont {P.~R.}\ \bibnamefont
  {Pasnoori}}, \bibinfo {author} {\bibfnamefont {N.}~\bibnamefont {Andrei}},\
  and\ \bibinfo {author} {\bibfnamefont {P.}~\bibnamefont {Azaria}},\
  }\bibfield  {title} {\bibinfo {title} {Edge modes in one-dimensional
  topological charge conserving spin-triplet superconductors: Exact results
  from {B}ethe ansatz},\ }\href {https://doi.org/10.1103/PhysRevB.102.214511}
  {\bibfield  {journal} {\bibinfo  {journal} {Phys. Rev. B}\ }\textbf {\bibinfo
  {volume} {102}},\ \bibinfo {pages} {214511} (\bibinfo {year}
  {2020})}\BibitemShut {NoStop}%
\bibitem [{\citenamefont {Kattel}\ \emph {et~al.}(2025)\citenamefont {Kattel},
  \citenamefont {Tang}, \citenamefont {Pixley},\ and\ \citenamefont
  {Andrei}}]{Andrei2025}%
  \BibitemOpen
  \bibfield  {author} {\bibinfo {author} {\bibfnamefont {P.}~\bibnamefont
  {Kattel}}, \bibinfo {author} {\bibfnamefont {Y.}~\bibnamefont {Tang}},
  \bibinfo {author} {\bibfnamefont {J.~H.}\ \bibnamefont {Pixley}},\ and\
  \bibinfo {author} {\bibfnamefont {N.}~\bibnamefont {Andrei}},\ }\bibfield
  {title} {\bibinfo {title} {Edge spin fractionalization in open
  one-dimensional spin-$s$ quantum antiferromagnets},\ }\href
  {https://doi.org/10.1103/PhysRevB.111.L220402} {\bibfield  {journal}
  {\bibinfo  {journal} {Phys. Rev. B}\ }\textbf {\bibinfo {volume} {111}},\
  \bibinfo {pages} {L220402} (\bibinfo {year} {2025})}\BibitemShut {NoStop}%
\bibitem [{\citenamefont {Pasnoori}\ and\ \citenamefont
  {Azaria}(2025)}]{Azaria2025}%
  \BibitemOpen
  \bibfield  {author} {\bibinfo {author} {\bibfnamefont {P.~R.}\ \bibnamefont
  {Pasnoori}}\ and\ \bibinfo {author} {\bibfnamefont {P.}~\bibnamefont
  {Azaria}},\ }\href {https://arxiv.org/abs/2506.19771} {\bibinfo {title}
  {Interplay between symmetry breaking and interactions in a symmetry protected
  topological phase}} (\bibinfo {year} {2025}),\ \Eprint
  {https://arxiv.org/abs/2506.19771} {arXiv:2506.19771 [hep-th]} \BibitemShut
  {NoStop}%
\bibitem [{\citenamefont {Montorsi}\ \emph {et~al.}(2017)\citenamefont
  {Montorsi}, \citenamefont {Dolcini}, \citenamefont {Iotti},\ and\
  \citenamefont {Rossi}}]{Montorsi2017}%
  \BibitemOpen
  \bibfield  {author} {\bibinfo {author} {\bibfnamefont {A.}~\bibnamefont
  {Montorsi}}, \bibinfo {author} {\bibfnamefont {F.}~\bibnamefont {Dolcini}},
  \bibinfo {author} {\bibfnamefont {R.~C.}\ \bibnamefont {Iotti}},\ and\
  \bibinfo {author} {\bibfnamefont {F.}~\bibnamefont {Rossi}},\ }\bibfield
  {title} {\bibinfo {title} {Symmetry-protected topological phases of
  one-dimensional interacting fermions with spin-charge separation},\ }\href
  {https://doi.org/10.1103/PhysRevB.95.245108} {\bibfield  {journal} {\bibinfo
  {journal} {Phys. Rev. B}\ }\textbf {\bibinfo {volume} {95}},\ \bibinfo
  {pages} {245108} (\bibinfo {year} {2017})}\BibitemShut {NoStop}%
\bibitem [{\citenamefont {Ludwig}(2016)}]{Ludwig2016}%
  \BibitemOpen
  \bibfield  {author} {\bibinfo {author} {\bibfnamefont {A.~W.~W.}\
  \bibnamefont {Ludwig}},\ }\bibfield  {title} {\bibinfo {title} {Topological
  phases: classification of topological insulators and superconductors of
  non-interacting fermions, and beyond},\ }\href
  {https://doi.org/10.1088/0031-8949/2015/T168/014001} {\bibfield  {journal}
  {\bibinfo  {journal} {Physica Scripta}\ }\textbf {\bibinfo {volume} {2016}},\
  \bibinfo {pages} {014001} (\bibinfo {year} {2016})}\BibitemShut {NoStop}%
\bibitem [{\citenamefont {Keselman}\ and\ \citenamefont
  {Berg}(2015)}]{Keselman2015}%
  \BibitemOpen
  \bibfield  {author} {\bibinfo {author} {\bibfnamefont {A.}~\bibnamefont
  {Keselman}}\ and\ \bibinfo {author} {\bibfnamefont {E.}~\bibnamefont
  {Berg}},\ }\bibfield  {title} {\bibinfo {title} {Gapless symmetry-protected
  topological phase of fermions in one dimension},\ }\href
  {https://doi.org/10.1103/PhysRevB.91.235309} {\bibfield  {journal} {\bibinfo
  {journal} {Phys. Rev. B}\ }\textbf {\bibinfo {volume} {91}},\ \bibinfo
  {pages} {235309} (\bibinfo {year} {2015})}\BibitemShut {NoStop}%
\bibitem [{\citenamefont {Kainaris}\ and\ \citenamefont
  {Carr}(2015)}]{Kainaris2015}%
  \BibitemOpen
  \bibfield  {author} {\bibinfo {author} {\bibfnamefont {N.}~\bibnamefont
  {Kainaris}}\ and\ \bibinfo {author} {\bibfnamefont {S.~T.}\ \bibnamefont
  {Carr}},\ }\bibfield  {title} {\bibinfo {title} {Emergent topological
  properties in interacting one-dimensional systems with spin-orbit coupling},\
  }\href {https://doi.org/10.1103/PhysRevB.92.035139} {\bibfield  {journal}
  {\bibinfo  {journal} {Phys. Rev. B}\ }\textbf {\bibinfo {volume} {92}},\
  \bibinfo {pages} {035139} (\bibinfo {year} {2015})}\BibitemShut {NoStop}%
\bibitem [{\citenamefont {Kainaris}\ \emph {et~al.}(2017)\citenamefont
  {Kainaris}, \citenamefont {Santos}, \citenamefont {Gutman},\ and\
  \citenamefont {Carr}}]{Kainaris2017}%
  \BibitemOpen
  \bibfield  {author} {\bibinfo {author} {\bibfnamefont {N.}~\bibnamefont
  {Kainaris}}, \bibinfo {author} {\bibfnamefont {R.~A.}\ \bibnamefont
  {Santos}}, \bibinfo {author} {\bibfnamefont {D.~B.}\ \bibnamefont {Gutman}},\
  and\ \bibinfo {author} {\bibfnamefont {S.~T.}\ \bibnamefont {Carr}},\
  }\bibfield  {title} {\bibinfo {title} {Interaction induced topological
  protection in one-dimensional conductors},\ }\href
  {https://doi.org/https://doi.org/10.1002/prop.201600054} {\bibfield
  {journal} {\bibinfo  {journal} {Fortschritte der Physik}\ }\textbf {\bibinfo
  {volume} {65}},\ \bibinfo {pages} {1600054} (\bibinfo {year}
  {2017})}\BibitemShut {NoStop}%
\bibitem [{\citenamefont {Thorngren}\ \emph {et~al.}(2021)\citenamefont
  {Thorngren}, \citenamefont {Vishwanath},\ and\ \citenamefont
  {Verresen}}]{Verresen2021}%
  \BibitemOpen
  \bibfield  {author} {\bibinfo {author} {\bibfnamefont {R.}~\bibnamefont
  {Thorngren}}, \bibinfo {author} {\bibfnamefont {A.}~\bibnamefont
  {Vishwanath}},\ and\ \bibinfo {author} {\bibfnamefont {R.}~\bibnamefont
  {Verresen}},\ }\bibfield  {title} {\bibinfo {title} {Intrinsically gapless
  topological phases},\ }\href {https://doi.org/10.1103/PhysRevB.104.075132}
  {\bibfield  {journal} {\bibinfo  {journal} {Phys. Rev. B}\ }\textbf {\bibinfo
  {volume} {104}},\ \bibinfo {pages} {075132} (\bibinfo {year}
  {2021})}\BibitemShut {NoStop}%
\bibitem [{\citenamefont {Santos}\ \emph {et~al.}(2016)\citenamefont {Santos},
  \citenamefont {Gutman},\ and\ \citenamefont {Carr}}]{Santos2016}%
  \BibitemOpen
  \bibfield  {author} {\bibinfo {author} {\bibfnamefont {R.~A.}\ \bibnamefont
  {Santos}}, \bibinfo {author} {\bibfnamefont {D.~B.}\ \bibnamefont {Gutman}},\
  and\ \bibinfo {author} {\bibfnamefont {S.~T.}\ \bibnamefont {Carr}},\
  }\bibfield  {title} {\bibinfo {title} {Phase diagram of two interacting
  helical states},\ }\href {https://doi.org/10.1103/PhysRevB.93.235436}
  {\bibfield  {journal} {\bibinfo  {journal} {Phys. Rev. B}\ }\textbf {\bibinfo
  {volume} {93}},\ \bibinfo {pages} {235436} (\bibinfo {year}
  {2016})}\BibitemShut {NoStop}%
\bibitem [{\citenamefont {Kainaris}\ \emph {et~al.}(2018)\citenamefont
  {Kainaris}, \citenamefont {Carr},\ and\ \citenamefont
  {Mirlin}}]{Kainaris2018}%
  \BibitemOpen
  \bibfield  {author} {\bibinfo {author} {\bibfnamefont {N.}~\bibnamefont
  {Kainaris}}, \bibinfo {author} {\bibfnamefont {S.~T.}\ \bibnamefont {Carr}},\
  and\ \bibinfo {author} {\bibfnamefont {A.~D.}\ \bibnamefont {Mirlin}},\
  }\bibfield  {title} {\bibinfo {title} {Transmission through a potential
  barrier in {L}uttinger liquids with a topological spin gap},\ }\href
  {https://doi.org/10.1103/PhysRevB.97.115107} {\bibfield  {journal} {\bibinfo
  {journal} {Phys. Rev. B}\ }\textbf {\bibinfo {volume} {97}},\ \bibinfo
  {pages} {115107} (\bibinfo {year} {2018})}\BibitemShut {NoStop}%
\bibitem [{\citenamefont {Matveeva}\ \emph {et~al.}(2025)\citenamefont
  {Matveeva}, \citenamefont {Gutman},\ and\ \citenamefont
  {Carr}}]{gapless_future2025}%
  \BibitemOpen
  \bibfield  {author} {\bibinfo {author} {\bibfnamefont {P.}~\bibnamefont
  {Matveeva}}, \bibinfo {author} {\bibfnamefont {D.}~\bibnamefont {Gutman}},\
  and\ \bibinfo {author} {\bibfnamefont {S.~T.}\ \bibnamefont {Carr}}}
  (\bibinfo {year} {2025}),\ \bibinfo {note} {in preparation}\BibitemShut
  {NoStop}%
\bibitem [{\citenamefont {Santos}\ \emph {et~al.}(2019)\citenamefont {Santos},
  \citenamefont {Gutman},\ and\ \citenamefont {Carr}}]{Santos2019}%
  \BibitemOpen
  \bibfield  {author} {\bibinfo {author} {\bibfnamefont {R.~A.}\ \bibnamefont
  {Santos}}, \bibinfo {author} {\bibfnamefont {D.~B.}\ \bibnamefont {Gutman}},\
  and\ \bibinfo {author} {\bibfnamefont {S.~T.}\ \bibnamefont {Carr}},\
  }\bibfield  {title} {\bibinfo {title} {Interplay between intrinsic and
  emergent topological protection on interacting helical modes},\ }\href
  {https://doi.org/10.1103/PhysRevB.99.075129} {\bibfield  {journal} {\bibinfo
  {journal} {Phys. Rev. B}\ }\textbf {\bibinfo {volume} {99}},\ \bibinfo
  {pages} {075129} (\bibinfo {year} {2019})}\BibitemShut {NoStop}%
\bibitem [{\citenamefont {Verresen}\ \emph {et~al.}(2017)\citenamefont
  {Verresen}, \citenamefont {Moessner},\ and\ \citenamefont
  {Pollmann}}]{Verresen2017}%
  \BibitemOpen
  \bibfield  {author} {\bibinfo {author} {\bibfnamefont {R.}~\bibnamefont
  {Verresen}}, \bibinfo {author} {\bibfnamefont {R.}~\bibnamefont {Moessner}},\
  and\ \bibinfo {author} {\bibfnamefont {F.}~\bibnamefont {Pollmann}},\
  }\bibfield  {title} {\bibinfo {title} {One-dimensional symmetry protected
  topological phases and their transitions},\ }\href
  {https://doi.org/10.1103/PhysRevB.96.165124} {\bibfield  {journal} {\bibinfo
  {journal} {Phys. Rev. B}\ }\textbf {\bibinfo {volume} {96}},\ \bibinfo
  {pages} {165124} (\bibinfo {year} {2017})}\BibitemShut {NoStop}%
\end{thebibliography}%
\end{document}